\documentclass{nature}

\usepackage{graphicx}
\usepackage{color}
\usepackage{soul}
\usepackage{amsmath}

\usepackage[normalem]{ulem}
	\usepackage{showlabels}

\newcommand{\bb}[1]{\mathbf{#1}}
\newcommand{\gen}[1]{\{#1\}}
\newcommand{\tl}[1]{\tilde{#1}}

\title{Generalized Maxwell projections for multi-mode network Photonics}

\author{M. Makarenko$^{1}$, A. Burguete-Lopez$^1$, F. Getman$^1$ \& A. Fratalocchi$^1$}

\begin{document}

\maketitle

\begin{affiliations}
\item PRIMALIGHT, Faculty of Electrical Engineering; Applied Mathematics and Computational Science, King Abdullah University of Science and Technology, Thuwal 23955-6900, Saudi Arabia 
\end{affiliations}


\begin{abstract}
The design of optical resonant systems for controlling light at the nanoscale is an exciting field of research in nanophotonics. While describing the dynamics of systems with few resonances is a relatively well understood problem, controlling the behavior of  systems with many overlapping states is considerably more difficult. In this work we use the theory of generalized operators to formulate an exact form of spatio-temporal coupled mode theory that retains  the simplicity of traditional coupled mode theory developed for optical waveguides. We developed a fast computational method that extracts all the characteristics of optical resonators, including the full density of states, the modes quality factors, the mode resonances and linewidths, by employing a single first principle simulation. This approach can facilitate the analytical and numerical study of complex dynamics arising from the interactions of many overlapping resonances in ensembles of resonators of any geometrical shape defined in materials with arbitrary responses. 
\end{abstract}

\section*{Introduction}
Dielectric optical nanoresonators are becoming an important platform for controlling light in nanoscale volumes of matter for many different applications~\cite{intro4,intro2,yuri4,intro1,intro3,maier2,res1,conti1,anten1,maier1}. The description of light-matter interactions in systems with two, or few, competing resonances is a relatively  understood subject~\cite{nonr1,nonr2,nonr3,nonr4,nonr5,pt1,pt2,bic1,yuri3,anapole,toroidal,yuri2,yuri5}. Controlling systems with many overlapping resonances, conversely, is more challenging. A main difficulty lies in the fact that resonator modes are usually derived from the solution of Maxwell equations with radiating boundary conditions and form a non orthogonal set, furnishing expressions that rapidly become difficult to manage, both analytically and in some cases also numerically, when the number of competing modes increases~\cite{Bohren1998}.\\
In the field of optical waveguides, the study of multi-modal systems is a mature and developed area of research in both linear and nonlinear settings~\cite{Yeh2008,agrawal2007nonlinear}. A significant contribution originates from the development in early days of exact theoretical frameworks that reduce Maxwell equations to simplified set of equations of motion, which furnish the building block to understanding complex hierarchical systems based on many interacting units~\cite{Kogelnik1975,Huang:94,1077767,nla.cat-vn1906292,tamir1979integrated}.\\         
In the field of optical resonators, an approximate form of this approach is available in time dependent coupled mode theory~\cite{haus,fan1}, which is routinely used in many applications to design of efficient broadband light energy trapping~\cite{Liu2013473,Liu2014f,Gomard:13}, the study of nonlinear dynamics~\cite{Shcherbakov2019}, and the engineering of photonic crystals and metamaterials~\cite{Joannopoulos:08:Book,Galinski2017d}. This theory derives equations of motion obtained under the condition of a total energy of the system $\mathcal E=\sum_m |a_m|^2$ expressed as the sum of independent terms $|a_m|^2$, each representing the energy of one resonant mode. The approximation originates from the lack of interacting contributions $a_ma^*_n$, which are necessary to account for the presence of non-orthogonal modes in the electromagnetic field expansion.\\
In this article, we aim at unifying these two areas by studying an exact form of spatio-temporal couple mode theory (STCMT), which retains the simplicity of time dependent equations developed for photonic resonators, and the exact nature of coupled mode equations studied for multi-mode optical waveguides. The approach is inspired by the Feshback operator splitting designed to study the spectral statistics of open quantum systems~\cite{PhysRevLett.89.083902,PhysRevA.67.013805,Liu2014f,PhysRevLett.114.043901,Totero_Gongora_2017,Cao_2005}, and here generalized to multiple projections ``spaces'' with the aid of different mathematics based on generalized functions~\cite{kanwal}.\\   
This formulation furnishes an intuitive description of Maxwell dynamics by providing an exact separation between propagating and resonant effects, within a simple set of exact equations that are particularly convenient for both analytical descriptions and numerical studies. We here illustrate fast numerical methods for calculating all the quantities of interest, ranging from the modes quality factors to the full density of states, from a single numerical simulation.   

\section*{Results}

\section*{Exact spatio-temporal coupled mode theory via generalized Maxwell projections}
\begin{figure*}
  \centerline{\includegraphics[width=1\textwidth]{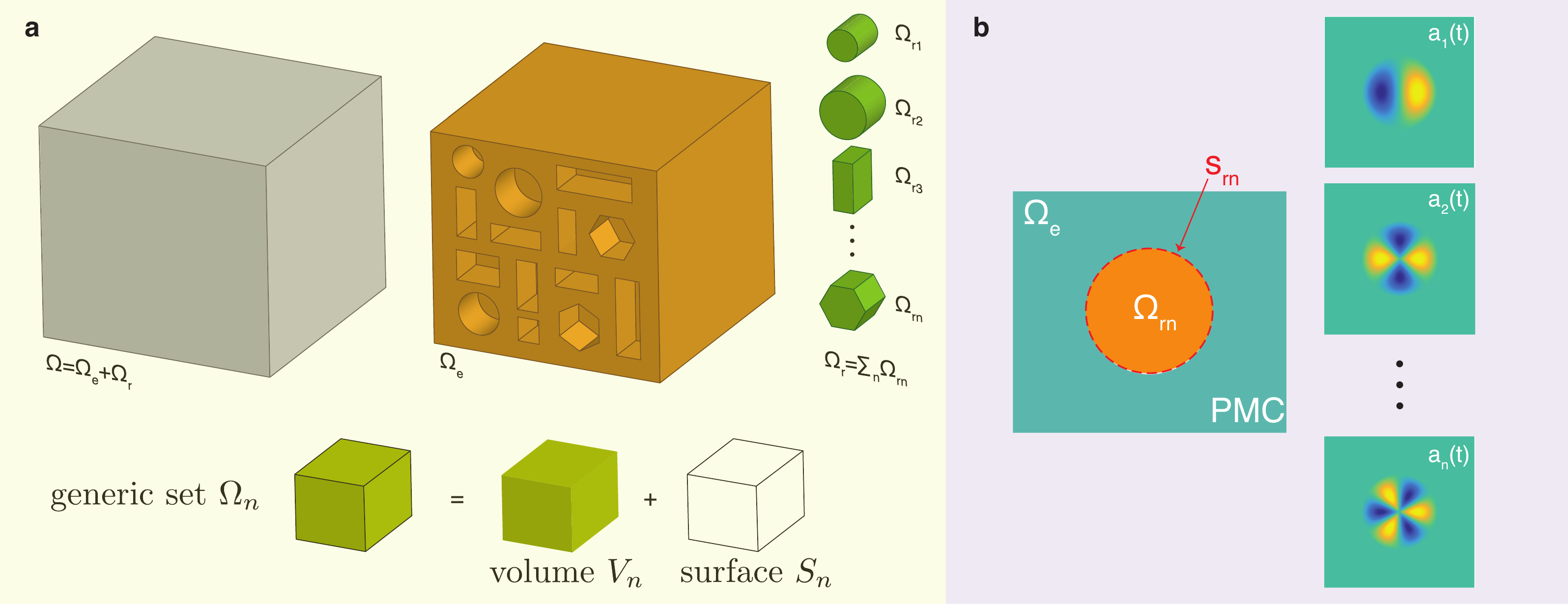}}
  \caption{\label{fig1}
    Example of space \(\Omega = \sum{\Omega_m}\) partitioning with different sets $\Omega_{m}$
characterized by elementary geometric structures. Panel b shows the equivalent dynamics inside $\Omega_{m}$, which contain a dielectric nanodisk resonator (orange area). Inside this space, $\Omega_e$ is seen as an ideal PEC material and the dynamics are decomposed as a series of electromagnetic modes of the nanodisk terminated by PEC boundary conditions.
   }
\end{figure*}

\begin{figure*}
  \centerline{\includegraphics[width=1\textwidth]{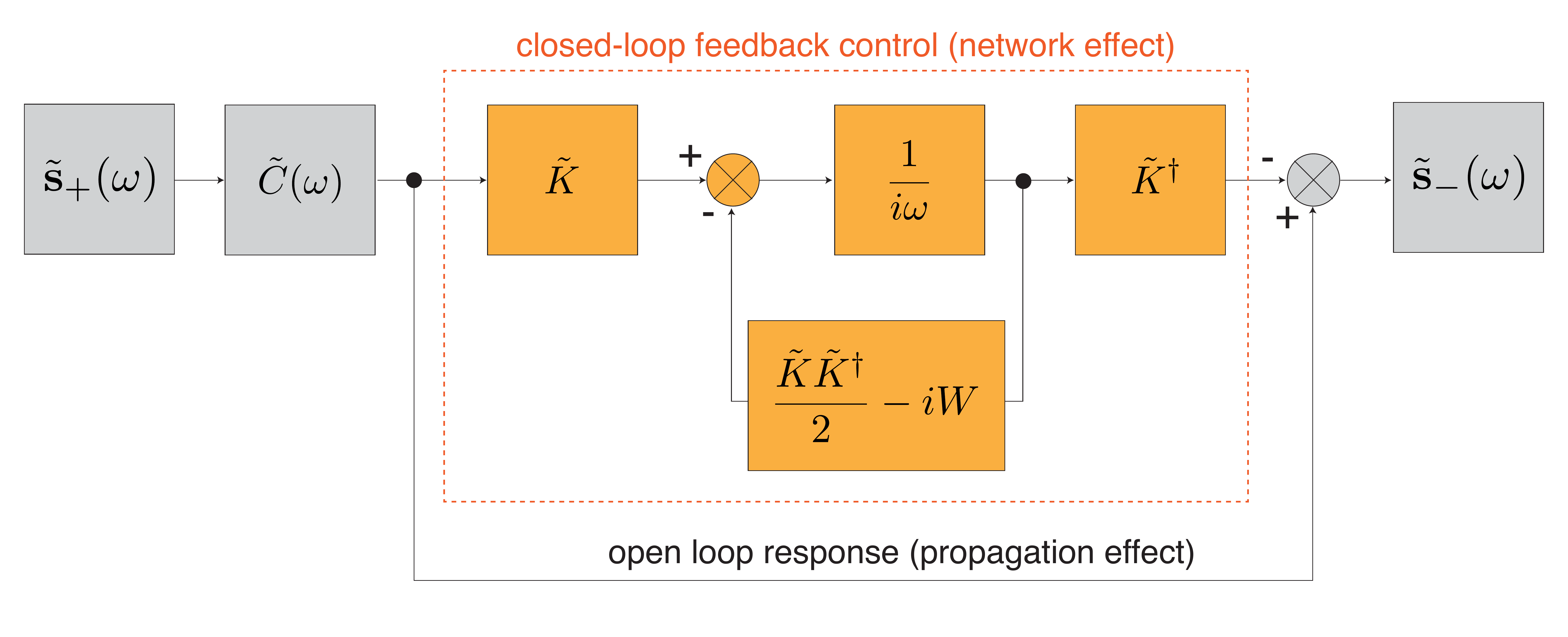}}
  \caption{\label{feedback}
  Block diagram representation of Eqs. \eqref{Eq.23}. The resulting electromagnetic dynamics are decomposed into propagation effects (open-loop) and resonance effects (blocks inside red-dashed area).
  }
\end{figure*}

The main idea of this approach is to divide the space $\Omega$ into a set of adjoined regions (Fig. \ref{fig1}), and formulate the dynamics of light evolution independently in each set through the use of orthogonal eigenmodes.\\
In the space decomposition $\Omega=\sum_n\Omega_{n}$, each region $\Omega_{n}$ is composed of an interior spatial volume $V_n$ and a boundary surface $S_n$ with a shape that is completely arbitrary. The union of all sets $\Omega_{n}$ containing at least one optical resonator inside their volume defines the resonator space $\Omega_r=\sum_n\Omega_{rn}$, while the remaining volume of matter builds up the external space $\Omega_e=\Omega-\Omega_r$.\\ 
In each set \(\Omega_{n},\) we can formulate Maxwell's equations by resorting to the theory of generalized functions~\cite{kanwal} and in particular by using the expression of the generalized differential operator $\nabla$, defined as follows:
\begin{equation}
\label{Eq.1}
\nabla=\{\nabla\}+\mathbf{n}_{S_{n}} \cdot \delta_{S_{n}},
\end{equation}
 $\{\nabla\}$ being the ordinary nabla operator evaluated at all the points inside the interior volume $V_n$, $\mathbf{n}_{S_{n}}$ the unit vector normal to the surface $S_n$, and $\delta_{S_n} = \delta(\mathbf{x}-\mathbf{x}_{S_n})$ a three-dimensional Dirac delta function centered on the surface $\mathbf{x}\in S_n$.
By substituting the expression of $\nabla$ from Eq. \eqref{Eq.1} into Maxwell equations, we obtain their generalized form:
\begin{equation}\label{Eq.2}
\left\{\begin{array}{l}{\{\nabla\} \times\left(\mathbf{E}_{n}+\mathbf{E}_{e}\right)+\delta_{S_n} \mathbf{n}_{S_n} \times\left(\mathbf{E}_{n}-\mathbf{E}_{e}\right)=-\mu(\mathbf{x}) \frac{\partial}{\partial t}\left(\mathbf{H}_{n}+\mathbf{H}_{e}\right)}, \\ {\{\nabla\} \times\left(\mathbf{H}_{n}+\mathbf{H}_{e}\right)+\delta_{S_n} \mathbf{n}_{S_n} \times\left(\mathbf{H}_{n}-\mathbf{H}_{e}\right)=\epsilon(\mathbf{x}) \frac{\partial}{\partial t}\left(\mathbf{E}_{n}+\mathbf{E}_{e}\right)}, \\ {\{\nabla\} \cdot\left(\mathbf{D}_{n}+\mathbf{D}_{e}\right)+\delta_{S_n} \mathbf{n}_{S_n} \cdot\left(\mathbf{D}_{n}-\mathbf{D}_{e}\right)=0}, \\ {\{\nabla\} \cdot\left(\mathbf{B}_{n}+\mathbf{B}_{e}\right)+\delta_{S_n} \mathbf{n}_{S_n} \cdot\left(\mathbf{B}_{n}-\mathbf{B}_{e}\right)=0.}\end{array}\right.
\end{equation}
In Eqs. \eqref{Eq.2}, subscripts $n$ and $e$ indicate fields defined in the $n$-th resonator region $\Omega_n$ and the external space $\Omega_e$, respectively, and \(\delta_{S_n}\mathbf{n}_{S_n}=\sum_{n} \mathbf{n}_{S_{n}}\cdot\delta\left(\mathbf{x}-\mathbf{x}_{S_{n}}\right)\) the singular contribution terms arising on the surfaces \(S_{n}\) separating the resonator space $\Omega_n$ from the environment.\\
We define each singular term $\delta\left(\mathbf{x}-\mathbf{x}_{S_{n}}\right)$ in Eqs. \eqref{Eq.2} as arising from the limiting condition $\varepsilon\rightarrow 0$ in which the surface $S_{n}$ is progressively approached from either the positive $(+)$ or negative $(-)$ side, indicated as $\delta\left(\mathbf{x}-\mathbf{x}_{S_{n}}\pm\varepsilon\right)$. This condition leads to equations that are mathematically exact in the limit of $\varepsilon\rightarrow 0$, and well defined for every value of $\varepsilon$. The choice of which direction $(\pm)$ to use to approach the surface $S_{n}$ is arbitrary and generates different boundary conditions for the equations defined in each region $\Omega_n$.\\
We here assume that the singular terms \(\delta_{S_{n+}} \mathbf{n}_{S_n} \times\left(\mathbf{E}_{n}-\mathbf{E}_{e}\right)\) and \(\delta_{S_{n+}}\mathbf{n}_{S_n} \cdot\left(\mathbf{B}_{n}-\mathbf{B}_{e}\right)\) are approached from the external space, with \(\delta_{S_{n+}}=\delta\left(\mathbf{x}-\mathbf{x}_{S}+\varepsilon\right)\), while the remaining singularities \(\delta_{S_{n-}}\mathbf{n}_{S_n} \times\left(\mathbf{H}_{n}-\mathbf{H}_{e}\right)\) and \(\delta_{S_{n-}}\mathbf{n}_{S_n} \cdot\left(\mathbf{D}_{n}-\mathbf{D}_{e}\right)\) are approached from within the resonator space, with \(\delta_{S_{n-}}=\delta\left(\mathbf{x}-\mathbf{x}_{S}-\varepsilon\right)\). This choice implies that, at every value of $\varepsilon$, the terms \(\delta_{S_{n+}} \mathbf{n}_{S_n} \times\left(\mathbf{E}_{n}-\mathbf{E}_{e}\right)\) and \(\delta_{S_{n+}}\mathbf{n}_{S_n} \cdot\left(\mathbf{B}_{n}-\mathbf{B}_{e}\right)\) are contained in the environment, while \(\delta_{S_{n-}}\mathbf{n}_{S_n} \times\left(\mathbf{H}_{n}-\mathbf{H}_{e}\right)\) and \(\delta_{S_{n-}}\mathbf{n}_{S_n} \cdot\left(\mathbf{D}_{n}-\mathbf{D}_{e}\right)\) are in the resonator space.\\
The choice splits Eqs. \eqref{Eq.2} into the following set, written for the $n$-th resonator $\Omega_n$ and the external space $\Omega_e$, respectively:
\begin{align}
\label{Eq.4}
&\Omega_n:\;\begin{cases}
\gen{\nabla}\times\bb{E}_r=-\mu(\bb x)\frac{\partial}{\partial t} \bb{H}_r,\\
\gen{\nabla}\times\bb{H}_r+\delta_{S_{n-}}\mathbf{n}_{S_n}\times(\bb{H}_r-\bb{H}_e)=\epsilon(\bb x)\frac{\partial}{\partial t}\bb{E}_r,\\
\gen{\nabla}\cdot\bb{D}_n+\delta_{S_{n-}}\mathbf{n}_{S_n}\cdot(\bb{D}_n-\bb{D}_e)=0,\\
\gen{\nabla}\cdot\bb{B}_n=0,\\
\end{cases}\nonumber\\
&\Omega_e:\;\begin{cases}
\gen{\nabla}\times\bb{E}_e+\delta_{S_{n+}}\mathbf{n}_{S_n}\times(\bb{E}_r-\bb{E}_e)=-\mu(\bb x)\frac{\partial}{\partial t} \bb{H}_e,\\
\gen{\nabla}\times\bb{H}_e=\epsilon(\bb x)\frac{\partial}{\partial t}\bb{E}_e,\\
\gen{\nabla}\cdot\bb{D}_e=0,\\
\gen{\nabla}\cdot\bb{B}_e+\delta_{S_{n+}}\mathbf{n}_{S_n}\cdot(\bb{B}_n-\bb{B}_e)=0.\\
\end{cases}
\end{align}
In Eqs. \eqref{Eq.4}, singular terms \(\delta_{S_{n+}} \mathbf{n}_{S_n} \times \mathbf{E}_{n}, \delta_{S_{n-}} \mathbf{n}_{S_n} \times \mathbf{H}_{n}, \delta_{S_{n+}} \mathbf{n}_{S_n} \cdot \mathbf{B}_{e}\), and \(\delta_{S_{n-}}\mathbf{n}_{S_n} \cdot \mathbf{D}_{n}\) represent the coupling of electromagnetic radiation at the surface of separation between the spaces $\Omega_n$ and $\Omega_e$. The remaining singular terms, conversely, define an appropriate set of boundary conditions.
In the case of zero electromagnetic field inside each resonator, $\mathbf E_n=0$ and light dynamics in the external space reduce to:
\begin{equation}\label{Eq.5}
\Omega_{e} : \left\{\begin{array}{l}{\{\nabla\} \times \mathbf{E}_{e}-\delta_{S_{n+}} \mathbf{n}_{S_n} \times \mathbf{E}_{e}=-\mu(\mathbf{x}) \frac{\partial}{\partial t} \mathbf{H}_{e}}, \\ {\{\nabla\} \times \mathbf{H}_{e}=\epsilon(\mathbf{x}) \frac{\partial}{\partial t} \mathbf{E}_{e}}, \\ {\{\nabla\} \cdot \mathbf{D}_{e}}=0, \\ {\{\nabla\} \cdot \mathbf{B}_{e}-\delta_{S_{n+}} \mathbf{n}_{S_n} \cdot \mathbf{B}_{e}=0}.\end{array}\right.
\end{equation}
In order for this system to be mathematically well defined, we need to impose the absence of any singular term. This implies setting $\delta_{S_{n+}} \mathbf{n}_{S_n} \times \mathbf{E}_{e}=\delta_{S_{n+}} \mathbf{n}_{S_n} \cdot \mathbf{B}_{e}=0$, which generates the following set of boundary conditions on $S_n$ in the limit $\varepsilon\rightarrow 0$:
\begin{equation}\label{Eq.6}
\left\{\begin{array}{l}{\mathbf{n}_{S_n} \times\left.\mathbf{E}_{e}\right|_{S_n}=0}, \\ {\mathbf{n}_{S_n} \cdot\left.\mathbf{H}_{e}\right|_{S_n}=0.}\end{array}\right.
\end{equation}
Equations \eqref{Eq.6} show that that the whole resonator space $\Omega_r=\sum_n\Omega_n$ is seen as a Perfect Electric Conductor (PEC) material from the external space.
Analogously, by imposing the absence of any singular terms in the dynamics of the resonator space when the external field is absent, we obtain the set of boundary conditions for $\Omega_e$:
\begin{equation}\label{Eq.7}
\left\{\begin{array}{l}{\mathbf{n}_{S_n} \times\left.\mathbf{H}_{n}\right|_{S_n}=0}, \\ {\mathbf{n}_{S_n} \cdot\left.\mathbf{D}_{n}\right|_{S_n}=0.}\end{array}\right.
\end{equation}
Equations \eqref{Eq.7} imply that the external space is seen from within each resonator region
$\Omega_{n}$ as a Perfect Magnetic Conductor (PMC) material. Boundary conditions \eqref{Eq.6}-\eqref{Eq.7} lead to the following final set of Maxwell equations:
\begin{equation}\label{Eq.8}
\begin{aligned}
\Omega_{n} : &\left\{\begin{array}{l}{\{\nabla\} \times \mathbf{E}_{n}=-\mu(\mathbf{x}) \frac{\partial}{\partial t} \mathbf{H}_{n}}, \\ {\{\nabla\} \times \mathbf{H}_{n}-\delta_{S_{n-}}\mathbf{n}_{S_n} \times \mathbf{H}_{e}=\epsilon_{0} \epsilon_{r}(\mathbf{x}) \frac{\partial}{\partial t} \mathbf{E}_{n}+\mathbf{J}_{\Delta}}, \\ {\{\nabla\} \cdot \mathbf{D}_{n}-\delta_{S_{n-}}\mathbf{n}_{S_n} \cdot \mathbf{D}_{e}=0}, \\ {\{\nabla\} \cdot \mathbf{B}_{n}=0}\end{array}\right. &  &
\mathrm{PMC} : \left\{\begin{array}{l}{\mathbf{n}_{S_n} \times\left.\mathbf{H}_{n}\right|_{S_n}=0}, \\ {\mathbf{n}_{S_n} \cdot\left.\mathbf{D}_{n}\right|_{S_n}=0},\end{array}\right. \\
\Omega_{e} : &\left\{\begin{array}{l}{\{\nabla\} \times \mathbf{E}_{e}+\delta_{S_{n+}} \mathbf{n}_{S_n} \times \mathbf{E}_{n}=-\mu(\mathbf{x}) \frac{\partial}{\partial t} \mathbf{H}_{e}}, \\ {\{\nabla\} \times \mathbf{H}_{e}=\epsilon(\mathbf{x}) \frac{\partial}{\partial t} \mathbf{E}_{e}}, \\ {\{\nabla\} \cdot \mathbf{D}_{e}}=0, \\ {\{\nabla\} \cdot \mathbf{B}_{e}+\delta_{S_{n+}} \mathbf{n}_{S_n} \cdot \mathbf{B}_{n}=0}\end{array}\right. &  & 
\operatorname{PEC} : \left\{\begin{array}{l}{\mathbf{n}_{S_n} \times\left.\mathbf{E}_{e}\right|_{S_n}=0}, \\ {\mathbf{n}_{S_n} \cdot\left.\mathbf{H}_{e}\right|_{S_n}=0}.\end{array}\right.
\end{aligned}
\end{equation}
In Eqs. \eqref{Eq.8} we have expanded the electric displacement \(\epsilon(\mathbf{x}) \frac{\partial}{\partial t} \mathbf{E}_{n}=\epsilon_{0} \epsilon_{n}(\mathbf{x}) \frac{\partial}{\partial t} \mathbf{E}_{n}+\mathbf{J}_{\Delta}\) into a linear contribution \(\epsilon_{n}(\mathbf{x}) \mathbf{E}_{n}\) and a generic source term \(\mathbf{J}_{\Delta}(\mathbf{x}, t)\) that keeps into account general types of effects, including dispersive effects,  amplification and nonlinear responses.
If we choose to approach the singular terms in Eqs. \eqref{Eq.4} in a different way, we obtain different combinations of ideal PEC/PMC boundary conditions.\\
The advantage of the splitting described by Eqs. \eqref{Eq.8} is to decompose the dynamics of light into different spatial regions terminated by ideal PEC/PMC boundary conditions, which allow to describe the evolution of the electromagnetic field with a complete eigenbasis of fully orthogonal modes.
In the resonator space, orthogonal modes are obtained from the eigenvalue problem of Maxwell equations, written inside each space $\Omega_n$:
\begin{align}
\label{Eq.9}
&\begin{bmatrix}
0 &\gen{\nabla}\times\\
-\gen{\nabla}\times &0
\end{bmatrix}\begin{bmatrix}
\bb E_{nm}\\
\bb H_{nm}
\end{bmatrix}
=i\omega_m
\begin{bmatrix}
\epsilon_0\epsilon_r(\bb x) &0\\
0 &\mu(\bb x)
\end{bmatrix}\begin{bmatrix}
\bb E_{nm}\\
\bb H_{nm}
\end{bmatrix},
\end{align}  
and terminated by PMC boundary conditions.
The operator \(\{\nabla\} \times\) is self-adjoint with PMC boundary conditions~\cite{jackson_classical_1999}. This implies that the resonator modes \(\mathbf{E}_{nm}\),
\(\mathbf{H}_{nm}\) are orthogonal, form a complete basis and possess a real frequency $\omega_m$. Mode orthogonality is calculated from the eigenvalue problem \eqref{Eq.9} using standard techniques~\cite{sakurai_napolitano_2017} and occurs through the following relationship:
\begin{equation}\label{Eq.10}
\frac{1}{2} \int_{V_{n}} d V_{n}\left[\mathbf{E}_{nm^{\prime}}^{*}, \mathbf{H}_{nm^{\prime}}^{*}\right] \left[ \begin{array}{cc}{\epsilon_{0} \epsilon_{r}(\mathbf{x})} & {0} \\ {0} & {\mu(\mathbf{x})}\end{array}\right] \left[ \begin{array}{c}{\mathbf{E}_{nm}} \\ {\mathbf{H}_{nm}}\end{array}\right]=\delta_{mm^{\prime}}.
\end{equation}
When \(m=m^{\prime}\), the integral expression in \eqref{Eq.10} represents the electromagnetic energy stored inside the resonator space \(\Omega_{n}\) by the $m$-th mode. Equation \eqref{Eq.10} is the counterpart of the orthogonality relation of guided modes in waveguides obtained via Pointing theorem (see, e.g., Eq. (2.2.52) of \cite{Kogelnik1975}) and offer the same formulation advantages: when the electromagnetic field inside the resonator is expanded in terms of resonator modes (\emph{c.c.} stands for complex conjugate):
\begin{equation}\label{Eq.11}
\mathbf{E}_{n}(\mathbf{x}, t)=\sum_{m=1}^{M} \frac{a_{m}(t)}{2} \mathbf{E}_{nm}(\mathbf{x})+c . c ., \quad \mathbf{H}_{n}(\mathbf{x}, t)=\sum_{m=1}^{M} \frac{a_{m}(t)}{2} \mathbf{H}_{nm}(\mathbf{x})+c . c . 
\end{equation}
The time averaged electromagnetic energy \(\langle\mathcal{E}(t)\rangle\) dissipated inside the resonator space for monochromatic excitation at frequency $\omega$ becomes simply expressed as the sum of the energy of each mode:
\begin{equation}\label{Eq.12}
\langle\mathcal{E}(t)\rangle=\int_{V_{n}} d V_{n}\left\langle\epsilon_{0} \epsilon_{r} \mathbf{E}_{n}^{2}+\mu \mathbf{H}_{n}^{2}\right\rangle=\sum_{m}\left|\tilde{a}_{m}(\omega)\right|^{2}=\tilde{\mathbf{a}}^{2},
\end{equation}
with \(a_{m}(t)=\tilde{a}(\omega)e^{i \omega t}\) and \(\tilde{\mathbf{a}}=\left[\tilde{a}_{1}(\omega), \ldots, \tilde{a}_{M}(\omega)\right]\))] defining the vector of amplitudes of the internal modes in the frequency domain. Equation \eqref{Eq.12} is the counterpart of the expression of the power in multi-mode waveguides, furnished by the squared sum of non-interacting terms (see, e.g., Eq. (2.2.56) of \cite{Kogelnik1975}). The expression \eqref{Eq.12} is only exact in the formulation of Eqs. \eqref{Eq.9} with PMC boundary conditions. \\ 
Analytic and closed form expressions of resonator modes and resonant frequencies in basic geometries are available from classical electrodynamics results of ideal metallic resonators~\cite{jackson_classical_1999}. As an example, for a single resonator space $\Omega_n$ characterized by a generic cuboid volume  with sides \(L_{x}, L_{y}, L_{z}\) along $(x,y,z)$ axis and filled with a dielectric material with refractive index $n(\mathbf{r})$, the frequencies $\omega_m$ of internal modes are:
\begin{equation}\label{Eq.13}
\begin{cases}{\omega=\omega_{l p q}=\frac{c}{n}\left|\mathbf{k}_{l p q}\right|}, \\ \mathbf{k}_{l p q}=\left(l\frac{\pi}{L_x}, p\frac{\pi}{L_y}, q\frac{\pi}{L_z}\right),\end{cases}
\end{equation}
with $l,p,q$ integers. The corresponding magnetic field distributions are then expressed as follows:
\begin{equation}\label{Eq.14}
\left\{\begin{array}{l}{H_{nx}(\mathbf{x})=A_{x} \cos \left(k_{x} x\right) \sin \left(k_{y} y\right) \sin \left(k_{z} z\right)} \\ {H_{ny}(\mathbf{x})=A_{y} \sin \left(k_{x} x\right) \cos \left(k_{y} y\right) \sin \left(k_{z} z\right)} \\ {H_{nz}(\mathbf{x})=A_{y} \sin \left(k_{x} x\right) \sin \left(k_{y} y\right) \cos \left(k_{z} z\right)}\end{array}\right.
\end{equation}
with $A_j$ being normalization constants.\\
External modes, existing in the outer region $\Omega_e$ are then expanded as a series of ingoing and outgoing scattered waves. As the radiation spectrum is typically continuous, it is convenient to carry out the mode expansion in the frequency domain \(\mathbf{E}_{e}(\mathbf{x}, t)=\tilde{\mathbf{E}}_{e}(\mathbf{x}, \omega) e^{i \omega t}, \mathbf{H}_{e}(\mathbf{x}, t)=\tilde{\mathbf{H}}_{e}(\mathbf{x}, \omega) e^{i \omega t}\):
\begin{equation}\label{Eq.15}
\begin{aligned}
\left\{\begin{array}{l}{\tilde{\mathbf{E}}_{e}=\sum_{h=1}^{H} \tilde{s}_{h+}(\omega) \mathbf{E}_{h+}+\tilde{s}_{h-}(\omega) \mathbf{E}_{h-}} \\ {\tilde{\mathbf{H}}_{e}=\sum_{h=1}^{H} \tilde{s}_{h+}(\omega) \mathbf{H}_{h+}+\tilde{s}_{h-}(\omega) \mathbf{H}_{h-}}\end{array}\right. & , &
\mathrm{PEC} : \left\{\begin{array}{l}{\mathbf{n}_{S_n} \times\left.\tilde{\mathbf{E}}_{e}\right|_{S_n}=0}, \\ {\mathbf{n}_{S_n} \cdot\left.\tilde{\mathbf{H}}_{e}\right|_{S_n}=0},\end{array}\right.
\end{aligned}
\end{equation}
with time varying amplitude coefficients \(s_{m \pm}(t)=\tilde{s}_{m \pm}(\omega) e^{i \omega t}\), which describe the time evolution of incoming \(\mathbf{E}_{m+}, \mathbf{H}_{m+}\) and outgoing \(\mathbf{E}_{m-}, \mathbf{H}_{m-}\) waves through $m=1,...,M$ different scattering channels. Traveling waves \(\mathbf{E}_{m \pm}, \mathbf{H}_{m \pm}\) depend in general on $\omega$ through their wavevector $\mathbf{k}$ as, e.g., in the case of plane waves \(e^{\pm i \mathbf{k} \cdot \mathbf{r}}\), spherical waves \(e^{\pm i k r / r}\) or other types of traveling waves in the free-space.\\ 
Following the same idea developed for the internal modes expansion in Eqs. \eqref{Eq.12}, we normalize radiating modes \(\tilde{\mathbf{E}}_{m \pm}\) and \(\tilde{\mathbf{H}}_{m \pm}\) through an observable quantity of physical interest. We here use the optical power $P$~\cite{tamir}, defined from the following integral when $h=h'$:
\begin{equation}\label{Eq.16}
\frac{1}{4} \int_{S} d S\left(\tilde{\mathbf{E}}_{h \pm} \times \tilde{\mathbf{H}}^*_{h' \pm}+\tilde{\mathbf{E}}_{h \pm}^{*} \times \tilde{\mathbf{H}}_{h' \pm}\right) \cdot \mathbf{n}_{S}=\delta_{hh^{\prime}},
\end{equation}
with $S=\sum_n S_n$ representing the union of all the surfaces of the resonator space $\Omega_r$. With the orthogonality condition \eqref{Eq.16}, the effective power $P$ flowing through $S$ assumes the expression:
\begin{equation}\label{Eq.17}
P=\frac{1}{2} \operatorname{Re}\left\{\int_{S} d S\left(\tilde{\mathbf{E}}_{e} \times \tilde{\mathbf{H}}_{e}^{*}\right) \cdot \mathbf{n}_{S}\right\}=\sum_{n}\left|\tilde{s}_{n+}(\omega)\right|^{2}-\sum_{n}\left|\tilde{s}_{n-}(\omega)\right|^{2}=\left|\tilde{\mathbf{s}}_{+}\right|^{2}-\left|\tilde{\mathbf{s}}_{-}\right|^{2}
\end{equation}
with \(\tilde{\mathbf{s}}_{\pm}=\left[\tilde{s}_{1 \pm}(\omega), \ldots, \tilde{s}_{M \pm}(\omega)\right]\)) defining the vector of incoming \((+)\) or outgoing \((-)\) waves.\\
The mode expansions carried out in Eqs. \eqref{Eq.12} and \eqref{Eq.15} reduce the time dynamics of Maxwell's equations to an exact set of spatio-temporal coupled mode equations, which relate the time evolution of the amplitudes of internal modes $a_m(t)$, with outgoing scattered waves $s_{h-}(t)$ for a given set of impinging sources $s_{h+}(t)$. The mode expansions in Eqs. \eqref{Eq.11} and Eqs. \eqref{Eq.15} express the corresponding spatial distribution of the field, providing a complete solution to the problem.\\
Coupled mode equations for the time varying coefficients are found in two different approaches. One technique is to expand the electromagnetic field with Eqs. \eqref{Eq.11} and \eqref{Eq.15}, substituting into Maxwell equations \eqref{Eq.8} and then projecting over each mode $a_m$ or $s_{h\pm}$ by using the orthogonality relations \eqref{Eq.10} and \eqref{Eq.16}. A second method is to exploit the linearity of Maxwell equations. We here employ a combination of both methods, starting from the latter.\\
In the external space $\Omega_e$, due to the absence of any source \(\mathbf{J}_{\Delta}=0\), Maxwell's equations are linear and the scattered modes $s_{h-}(t)$ follow a linear evolution as a function of modes $a_m(t)$ and impinging fields $s_{h+}(t)$. The time dynamics of the scattered field $\mathbf s_{-}(\omega)$ in the frequency domain is then expressed as a linear superposition of $\tilde{\mathbf{a}}(\omega)$ and $\tilde{\mathbf{s_+}}(\omega)$:
\begin{equation}\label{Eq.19}
\tilde{\mathbf{s}}_{-}(\omega)=\tilde{D}(\omega) \cdot \tilde{\mathbf{a}}(\omega)+\tilde{C}(\omega) \cdot \tilde{\mathbf{s}}_{+}(\omega),
\end{equation}
with \(\tilde{D}(\omega), \tilde{C}(\omega)\) being linear matrices. To write the equations describing the dynamics of \(\mathbf{a}(t)\), we first consider in the case of linear materials with \(\mathbf{J}_{\Delta}=0\). In this limit, Maxwell equations inside the resonator space $\Omega_n$ are also linear, and the dynamics of \(\mathbf{a}(t)\) follow from the most general form of linear time evolution of modes $a_m(t)$ with input sources corresponding to impinging waves $\mathbf s_+$ :
\begin{equation}\label{Eq.20}
\dot{\mathbf{a}}(t)=\int_{t_{0}}^{t} \mathrm{d} t^{\prime}\left[H\left(t-t^{\prime}\right) \cdot \mathbf{a}\left(t^{\prime}\right)+K\left(t-t^{\prime}\right) \cdot \mathbf{s}_{+}\left(t^{\prime}\right)\right].
\end{equation}
In Eq. \eqref{Eq.20}, \(\dot{\mathbf{a}}=\mathrm{d} \mathbf{a} / \mathrm{d} \mathrm{t}\), and \(H(t), K(t)\) are linear matrices with Fourier pairs \(\tilde{H}(\omega), \tilde{K}(\omega)\) in the frequency space. As any general matrix with no predefined symmetry, the matrix $H$ is decomposed as $H=iW+\Gamma$, with a Skew-Hermitian matrix $iW$ and an Hermitian matrix $\Gamma$. Without loss of generality, we can assume that the matrix $W$ is diagonal. If not, due to the Hermitian nature of $W$, we can always diagonalise $W$ by a unitary matrix and project $a_m$ into a new orthogonal basis that will preserve the energy relation \eqref{Eq.12}.\\
Matrices \(\tilde{H}, \tilde{K}, \tilde{C}, \tilde{D}\)  are not independent, as the dynamics resulting from \eqref{Eq.19}-\eqref{Eq.20} has to satisfy energy conservation:
\begin{equation}\label{Eq.21}
\left\langle\frac{d \mathcal{E}(t)}{d t}\right\rangle=\left|\tilde{\mathbf{s}}_{+}(\omega)\right|^{2}-\left|\tilde{\mathbf{s}}_{-}(\omega)\right|^{2}.
\end{equation}
By substituting the coupled mode equations \eqref{Eq.19}-\eqref{Eq.20} into Eq. \eqref{Eq.21}, we obtain the following self-consistency relations:
\begin{equation}\label{Eq.22}
\left\{\begin{array}{l}{\tilde{C}(\omega)^{\dagger} \tilde{C}(\omega)=1} \\ {2 \tilde{\Gamma}(\omega)=\tilde{K}(\omega) \tilde{K}^{\dagger}(\omega)} \\ {\tilde{D}(\omega)=-\tilde{C}(\omega) \tilde{K}^{\dagger}(\omega)}\end{array}\right.
\end{equation}
where $\widetilde{\Gamma}(\omega)$ is the Fourier transform of $\Gamma$. Equations \eqref{Eq.21} are a particular form of the fluctuation dissipation theorem \cite{kubo} applied for Maxwell equations. The conditions imposed by Eqs. \eqref{Eq.22} solve Eqs. \eqref{Eq.19}-\eqref{Eq.20} in the frequency domain:
\begin{equation}\label{Eq.23}
\begin{cases}
\tilde{\bb a}(\omega)=\frac{\tilde K}{i\left(\omega-W\right)+\frac{\tilde K \tilde{K}^\dag}{2}}\tl{\bb s}_+,\\
\tilde{\bb s}_-(\omega)=\tilde C\left(\tilde{\bb s}_+-\tl{K}^\dag\cdot\tilde{\bb a}\right),
\end{cases}
\end{equation}
with $1 / \tilde{X}$ being shorthand notation for the inverse matrix $\tilde{X}^{-1}$. Equations \eqref{Eq.23} are similar to the time dependent coupled mode equations written in the frequency domain and originally introduced in~\cite{haus,1337032}. However, there are also differences. In the traditional set~\cite{haus,1337032}, all the linear matrices $C$, $K$, $\Gamma$ are frequency independent and   $a_m(t)$ are the amplitudes of traditional electromagnetic modes with radiating boundary conditions.\\ 
 Figure \ref{feedback} shows a block diagram representation of Eqs. \eqref{Eq.23}. In the absence of any resonance, $\tl{\bb a}=0$ and the system output is characterized by the open loop response $\tl{\bb s}_-=\tl{C}(\omega)\cdot\tl{\bb s}_+$. This is the contribution that arises from purely propagation effects and away from any resonant light-matter interaction.
When $\tl{\bb a}\neq 0$, the system response is characterized by a second term represented by the closed-loop feedback unit of Fig. \ref{feedback}, which forms the contribution of resonances.\\
Equations \eqref{Eq.23} and Fig. \ref{feedback} show that the dynamics of Maxwell's equations depend only on three independent matrices: $\tilde{K}$,  $\tilde{C}$, and $W$. The physical meaning of these matrices and their expressions is analyzed next. When $\tilde{K}=0$ in Eqs. \eqref{Eq.23}, the $n$-th resonator $\Omega_n$ and the external $\Omega_e$ space are uncoupled, and the dynamics of \eqref{Eq.23} reduce to:
\begin{equation}\label{Eq.25}
\left\{\begin{array}{l}{i(\omega-W) \cdot \tilde{\mathbf{a}}(\omega)=0}, \\ {\tilde{\mathbf{s}}_{-}(\omega)=\tilde{C}(\omega) \cdot \tilde{\mathbf{s}}_{+}(\omega)}\end{array}\right. .
\end{equation}
The first equation describes the undamped motion of the internal modes $
a_{m}(t)
$ at frequencies $\omega_m$ arising from  the diagonal elements of the (resonances) matrix $W_{mm'}=\omega_m\delta_{mm'}$ (Fig. \ref{fig1}b). This dynamics represent free oscillations of non- interacting, orthogonal modes of the resonator space solution of Eq. \eqref{Eq.19}.\\ 
In the external space $\Omega_{e}$, as obtained from the second of Eqs. \eqref{Eq.25}, light dynamics reduces to a scattering
process of input sources $\mathbf{s_+}$ impinging on
the PMC material defined in each $\Omega_{n}$ representing a resonator. The matrix $\tilde C(\omega)$ is the unitary scattering matrix describing this process. The off-diagonal terms $
\tilde{C}_{m n}=\frac{\tilde{s}_{m-}}{\tilde{s}_{n+}}
$ of the matrix $\tilde C(\omega)$ represent the scattering of energy from incoming waves 
 $\tilde{s}_{m+}$ into outgoing radiation on different modes 
$\tilde{s}_{m-}$, while the diagonal terms 
$\tilde{C}_{n n}=\frac{\tilde{s}_{n-}}{\tilde{s}_{n+}}$ are the reflection coefficients of incoming waves \(\tilde{s}_{n+}\) into contributions propagating in the same mode but in
opposite directions \(\tilde{s}_{n-}\).\\
The scattering matrix can be expressed in exponential form \(\tilde{C}(\omega)=e^{i \phi(\omega)}\), with $\phi$ being $M \times M$ matrix. It is always possible to obtain an input output representation of the dynamics where the equivalent scattering matrix is the identity matrix. This is accomplished by defining a new vector of outgoing scattered waves as follows:
\begin{equation}\label{Eq.27}
\tilde{\mathbf{s}}_{-}^{\prime}=e^{-i \phi} \cdot \tilde{\mathbf{s}}_{-}, \quad \quad \tilde{\mathbf{s}}_{-}^{\prime}=\mathbf{1} \cdot\left(\tilde{\mathbf{s}}_{+}-\tilde{K}^{\dagger} \cdot \tilde{\mathbf{a}}\right).
\end{equation}
Equation \eqref{Eq.27} does not alter the space partitioning \(\Omega_{n}, \Omega_{e}\), nor the mode \(a_{n}(t)\) evolution. The transformation \eqref{Eq.27} defines a new set of scattering modes via \eqref{Eq.15} that diagonalise the scattering matrix and, as such, provide only reflections for each input channel excited in the dynamics. An example of this representation is furnished in the next section.\\
When the spaces $\Omega_e$ and $\Omega_n$ interact with nonzero couplings $\tilde K$, electromagnetic energy flows from the cavity region to $\Omega_e$, and viceversa. The coupling matrix \(\tilde{K}\) has in general a small, or weak dependence on the frequency $\omega$. This condition, known as Markov approximation of open quantum systems~\cite{doi:10.1063/1.531919,PhysRevA.67.013805}, is here discussed from the generalized Maxwell Eqs.~\eqref{Eq.8}.\\
By substituting the field expansion \eqref{Eq.15} in \eqref{Eq.8} and by projecting over each traveling mode, we obtain the coupling coefficient elements:
\begin{equation}\label{Eq.28}
\tilde{K}_{mn} \propto \int_{S} d S\left(\mathbf{n}_{S_n} \times \mathbf{E}_{n}\right) \cdot \mathbf{H}_{m \pm}^{*},
\end{equation}
with $\mathbf{E}_{n}$ being the field inside the resonator space and $\mathbf{H}_{m \pm}^{*}$ the scattering modes in the $m$-th channel. 
The time average of the integral gives a nonzero contribution 
only when the fields  $\mathbf{H}_{m \pm}^{*}$ and $\mathbf{E}_{n}$ are 
in phase
.\\
We discuss this condition with an illustrative example, derived in the case of a continuous external spectrum of plane waves \(e^{\pm i \mathbf{k} \cdot \mathbf{r}}\) interacting with cuboid resonator structures $\Omega_n$ with resonant wavevectors \(\mathbf{k}_{l p q}\) represented Eqs. \eqref{Eq.13} For a general frequency value $\omega=c\mathbf{k}$ that is not resonant with any internal resonance \(\mathbf{k} \neq \mathbf{k}_{l p q}\), the integral \eqref{Eq.28} is characterized by oscillatory terms of the type \(e^{i\left(\mathbf{k}-\mathbf{k}_{l p q}\right) \cdot \mathbf{r}}\), which integrated through $S$ do not furnish  contribution. In all of these cases, the coupling matrix \(\tilde{K}\) becomes frequency independent.\\
In the situation where \(\mathbf{J}_{\Delta} \neq 0\), the contribution of \(\mathbf{J}_{\Delta}\) to the dynamics is calculated by projecting Eqs. \eqref{Eq.8} over the internal eigenmodes of the resonator space, thus obtaining an additional source term in the dynamics of the internal modes:
\begin{equation}\label{Eq.29}
\dot{\mathbf{a}}(t)=\int_{t_{0}}^{t} \mathrm{d} t^{\prime}\left[H\left(t-t^{\prime}\right) \cdot \mathbf{a}\left(t^{\prime}\right)+K\left(t-t^{\prime}\right) \cdot \mathbf{s}_{+}\left(t^{\prime}\right)\right]-\mathbf{j}_{\Delta}(t),
\end{equation}
with \(\mathbf{j}_{\Delta}=\left[j_{\Delta 1}(t), \ldots, j_{\Delta m}(t)\right]\) being a vector of projected source terms with general contribution:
\begin{equation}\label{Eq.30}
 j_{\Delta m}(t)=\frac{\epsilon_0}{4} \int_{V_{n}} d V_{n} \epsilon_r(\mathbf{x}) \cdot\mathbf{E}_{m}^{*}(\mathbf{x}) \cdot \mathbf{J}_{\Delta}(\mathbf{x}, t),
\end{equation}
Equations \eqref{Eq.29}-\eqref{Eq.30} model the exact dynamics of light matter interactions in multimodal material structures with arbitrary defined linear and nonlinear responses.


\subsection{Quantities that can be calculated with this approach}
\begin{figure*}
  \centerline{\includegraphics[width=1\textwidth]{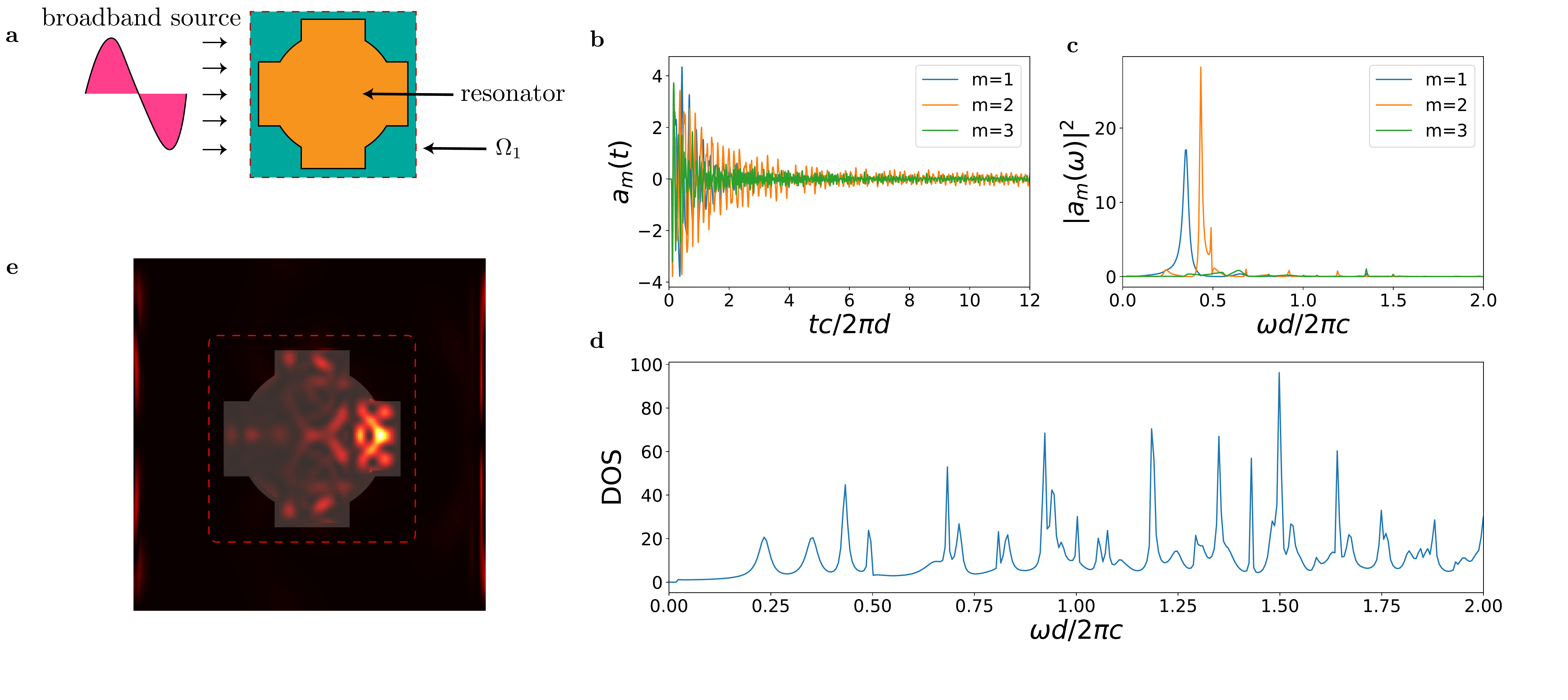}}
  \caption{\label{Panel2}
    Illustration of numerical technique to calculate DOS from a single FDTD simulation. Panel a shows the setup geometry, with an optical resonant structure (orange area) defined inside a cubic resonator space $\Omega_1$ illuminated by a broadband pulse. Panel b shows the time evolution of the resonant modes $a_m(t)$. Panel c shows their power density spectra. The corresponding density of states obtained by summing up all the modes contributions is shown in panel d. Panel e shows the electromagnetic energy density obtained from FDTD simulations at a certain time step. The red dotted rectangle represents a boundary of the cubic resonator space $\Omega_1$.
   }

\end{figure*}

\paragraph{Density of States.}
One of the most important quantities of a resonant system is the density of states (DOS), which is defined \cite{economou} as follows: 
\begin{equation}\label{Eq.31}
\operatorname{DOS}(\omega)=\sum_{n} \delta\left(\omega-\omega_n\right),
\end{equation}
and providing the number of eigenstates $\omega_n$ in the frequency interval $[\omega, \omega+d\omega]$.
The most common technique for calculating the DOS of an optical structure characterized by complex geometries and dispersive effects is to extract it numerically via, e.g., finite difference time domain (FDTD) simulations, by injecting an impulsive point-dipole source $\mathbf{S}_p$ in the electric $\mathbf{E}$ or magnetic $\mathbf{H}$ field such as:
\begin{equation}
    \label{pso0}
    \mathbf{S}_p(\mathbf{x},t)=\hat{\mathbf{e}}_l \delta\left(\mathbf x-\mathbf x_{0}\right) p(t),
\end{equation}
with $\hat{\mathbf{e}}_l$ ($l=x,y,z$) being a unit vector along one coordinate axis, $\mathbf{x}_0$ the coordinates of a point inside the material whose DOS is to be computed, and $p(t)=\delta(t-t_0)$ a short time pulse with broadband spectrum.\\
The local density of states (LDOS) measured at the point $\mathbf{x}_0$ and polarization $l$ is obtained from the power density spectrum of the electric or magnetic field measured at $\mathbf{x}_0$ via the following relation~\cite{taflove}:    
\begin{equation}\label{Eq.34}
\operatorname{LDOS}_{\ell}(\mathbf{x}_0, \omega)=\sum_{n} \delta\left(\omega-\omega_n\right) \varepsilon(\mathbf{x}_0)\left|E_{\ell_n}(\mathbf{x}_0)\right|^{2},
\end{equation}
Once the LDOS is calculated, the DOS is obtained by integrating over the volume of the material $V$ and by summing up the contributions arising from different polarizations:
\begin{equation}\label{Eq.33}
\operatorname{DOS}(\omega)=\sum_{l=x,y,z} \int \operatorname{LDOS}_{l}(\mathbf x, \omega)dV.
\end{equation}
Equation \eqref{Eq.33} is rarely employed in practice due to the requirement to perform a large number of simulations, in principle one for each point $\mathbf{x}_0$ and polarization considered.\\
The theory developed in the previous section allows for a fast calculation, which can furnish the complete DOS with just a single FDTD simulation.\\
By substituting the expression of the electromagnetic field inside each resonator region $\Omega_n$, given by Eq. \eqref{Eq.11}, into Eq. \eqref{Eq.33} and by integrating over the volume $V_n$ defined by the resonator region, we obtain the DOS corresponding to the resonator region $\Omega_n$ from the  sum of the power density spectra of the internal modes:
\begin{align}
    \label{dos11}
        \mathrm{DOS}(\omega)= &\epsilon_0\int dV_n\epsilon_r(\mathbf{x})\left\lvert\tilde{\mathbf{E}}_n(\mathbf{x},\omega)\right\rvert^2=\epsilon_0\nonumber\\
        &\times\sum_{mm'}\tilde{a}_m(\omega)\tilde{a}_{m'}(\omega)\int \mathbf{E}^*_{nm'}(\mathbf{x})\epsilon_r(\mathbf{x})\mathbf{E}_{nm}(\mathbf{x},\omega)dV_n=\sum_m\lvert \tilde{a}_m(\omega)\rvert^2,
\end{align}
where the last step is obtained through the orthogonality relations \eqref{Eq.10}.
Equation \eqref{dos11} can be evaluated with a single FDTD computation, by injecting a broadband, three dimensional dipole source $S_p(\mathbf{x},t)$ centered at any point outside $\Omega_n$, and then measuring the time evolution of the internal modes $a_m(t)$ of the resonator space $\Omega_n$ by projecting the electromagnetic field $\mathbf{E}_n$ or $\mathbf{H}_n$ over the corresponding eigenmode $\mathbf{E}_{nm}$ or $\mathbf{H}_{nm}$ via the orthogonality relations \eqref{Eq.10}:
\begin{equation}
    \label{mode0}
    a_m(t)=\frac{\epsilon_0}{2}\int \mathbf{E}^*_{nm}(\mathbf{x})\epsilon_r(\mathbf{x})\mathbf{E}_{n}(\mathbf{x},t)dV_n.
\end{equation}
Equation \eqref{mode0} can be evaluated during the FDTD simulation, as the electric field $\mathbf{E}_n(\mathbf{x},t)$ is available at each time $t$, and modal distributions $\mathbf{E}_{nm}(\mathbf{x})$ are easily calculated for resonators terminated by ideal  PEC/PMC boundary conditions by eigenvalue solvers~\cite{LoggMardalEtAl2012}. Once the distribution of $a_m(t)$ is known, the DOS is directly computed with Eq. \eqref{dos11}.\\ 
In the calculation of the DOS in the resonator space $\Omega_n$, it is also possible to use any orthogonal set of internal modes that results from the eigensolution of Eq. \eqref{Eq.9} with PEC/PMC boundaries and arbitrary material properties $\epsilon(\mathbf{x})$ and $\mu(\mathbf{x})$. \\
This result is demonstrated from the orthogonality and completeness of the modes. Let Eq. \eqref{Eq.11} describe the electromagnetic field $\mathbf{E}_n$, $\mathbf{H}_n$ inside a resonator structure $\Omega_n$, written in compact form as follows:
\begin{equation}
    \label{newm00}
    \begin{bmatrix}
    \mathbf{E}_n(\mathbf{x},t)\\
    \mathbf{H}_n(\mathbf{x},t)
    \end{bmatrix}=\sum_m\frac{a_{m}(t)}{2}\begin{bmatrix}
    \mathbf{E}_{nm}(\mathbf{x})\\
    \mathbf{H}_{nm}(\mathbf{x})
    \end{bmatrix}+c.c,
\end{equation}
We can then expand the same electromagnetic field by using a different set of eigenmodes pertaining to a material in $\Omega_n$ with permittivity $\epsilon^{(0)}(\mathbf{x})$ and permeability $\mu^{(0)}(\mathbf{x})$ calculated by using the same set of PEC/PMC boundary condition.\\
By projecting Eq. \eqref{newm00} on the new set of modes $\mathbf{E}^{(0)}_{nm}(\mathbf{x})$, $\mathbf{H}^{(0)}_{nm}(\mathbf{x})$, we obtain a new set of time varying amplitudes $b_m(t)$ related to $a_m(t)$ as follows:
\begin{align}
    \label{newm0}
    &a_m(t)=\sum_{m'}C_{mm'}b_{m'}(t), \nonumber\\
    &C_{mm'}=\frac{1}{2} \int_{V_{n}} d V_{n}\left[\mathbf{E}_{nm}^{*}, \mathbf{H}_{nm}^{*}\right] \left[ \begin{array}{cc}{\epsilon_{0} \epsilon_{r}(\mathbf{x})} & {0} \\ {0} & {\mu(\mathbf{x})}\end{array}\right] \left[ \begin{array}{c}{\mathbf{E}^{(0)}_{nm}} \\ {\mathbf{H}^{(0)}_{nm}}\end{array}\right].
\end{align}
Correspondingly, the density of states becomes:
\begin{equation}
    \label{newdos}
    \mathrm{DOS}=\sum_m\lvert \tilde a_m\rvert^2=\sum_{m,m',m''}C_{mm'}^*C_{mm''}\tilde b^*_{m'}\tilde b_{m''}=\sum_{m'm''}\delta_{m'm''}\tilde b^*_{m'}\tilde b_{m''}=\sum_{m'}|\tilde b_{m'}|^2,
\end{equation}
where the third equality stems from the completeness and orthogonality of the  modes, as the reader can verify from \eqref{newm0}. Equation \eqref{newm0} implies that the calculation of the DOS does not rely on the particular set of modes used, as long as they are computed with PEC/PMC boundary conditions.\\
A particularly convenient choice in the case of dielectric structures are modes of completely filled cuboid resonator structures with constants $\epsilon^{(0)}(\mathbf{x})=\epsilon_0\epsilon_r$ and $\mu^{(0)}=\mu_0$, which are analytically expressed by Eqs. \eqref{Eq.13}-\eqref{Eq.14} via simple trigonometric formulas. Other possible simple choices are represented by spherical or cylindrical $\Omega_n$ spaces characterized by analytic combination of Bessel functions. 
When using these equivalent modes expansions, the resonance matrix $W$ appearing in Eqs. \eqref{Eq.25} is in general not diagonal, due to Eqs. \eqref{newm0} that project $W$ into a matrix of full rank.\\
Figure \ref{Panel2} summarizes this procedure with an example of DOS calculation. We consider a resonator schematically illustrated in Fig. \ref{Panel2}a (orange area). We partition the space by using a cuboid resonator region $\Omega_1$ (Fig. \ref{Panel2}a green area). We illuminate the structure by a single broadband pulse source (Fig. \ref{Panel2} input pulse) and calculate the amplitudes $a_m(t)$ of the internal modes by Eq. \eqref{mode0}. Figure \ref{Panel2}b-c shows the time evolution of $a_m(t)$ and $|\tilde{a}(\omega)|^2$ for the first modes $m=1,2,3$.
The resulting DOS, calculated from Eq. \eqref{dos11}, is reported in Fig. \ref{Panel2}d. The energy density distribution from FDTD simulations at the time $t\cdot\frac{c}{2\pi d}=2$ shown on Fig. \ref{Panel2}e with red dotted rectangle representing a cuboid resonator region $\Omega_1$.  

\subsection{Photonic resonance networks}
\begin{figure*}
  \centerline{\includegraphics[width=0.75\textwidth]{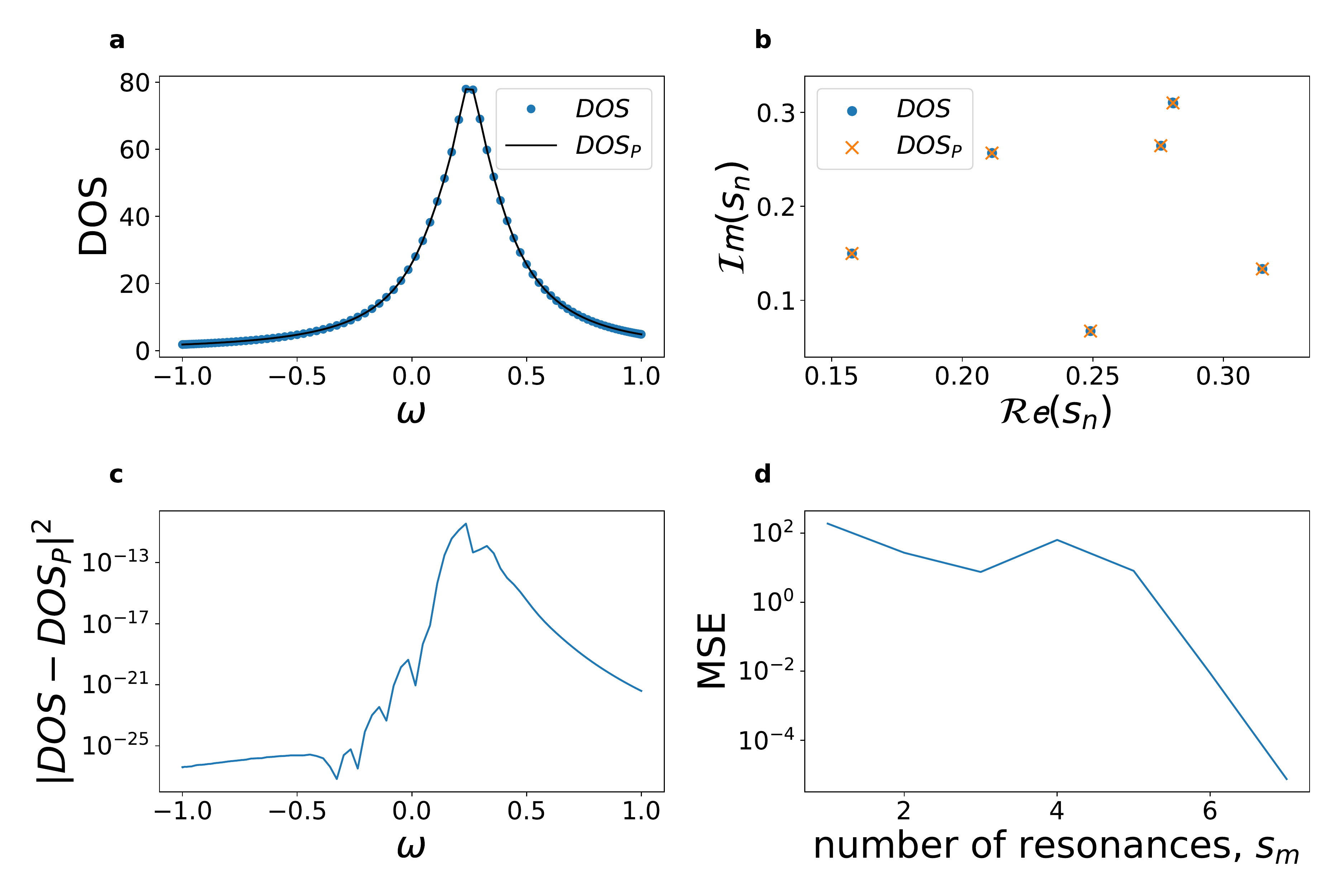}}
  \caption{\label{fitex} Stable rational function fitting algorithm. (a) Single resonant linewidth (circle markers) generated by the sum of seven overlapping resonances $s_n$ characterized by random frequencies and damping factors (b, circle marker). (c) Residual error between the original DOS and the predicted DOS ($\mathrm{DOS_P}$) from the fitting algorithm. (d) Mean square error (MSE) of the fitting procedure for increasing number of resonances. The numerically computed distribution of resonances $s_m$ is reported in panel b (cross markers) and the corespondent $\mathrm{DOS_P}$ in panel a (solid line).
   }
\end{figure*}

A particularly important quantity in the analysis of resonant systems is the mode quality factor $Q=\frac{\omega_0\tau}{2}$, defined as the product between the mode frequency $\omega$ and the mode decay rate $\tau$, and describing the ability of the mode to trap and release electromagnetic energy~\cite{haus}. Traditionally, the evaluation of the $Q$ factor requires to selectively excite each mode and compute $\omega$ and $\tau$ from the mode decaying rates in time, or from the mode frequency linewidth $\Delta$ in the LDOS. This approach assumes non-interacting resonances and cannot be directly applied in the general case of overlapping resonant states in the spectrum.\\
The theory developed via the generalized Maxwell's equations allows  to precisely evaluate the $Q$ factor of all the resonances of the system from a single simulation in the general case of interacting modes. To illustrate the calculation, we begin by solving Eq. \eqref{Eq.20} in the Markov limit where the input frequency $\omega$ is away from a resonant frequency of the system:
\begin{equation}
\label{exp01}
\mathbf{a}(t)=e^{Ht}\mathbf{a}_0+\int_{0}^{t} e^{H(t-t')}K\mathbf{s_+}(t') dt',
\end{equation}
with $\mathbf{a}_0=\mathbf{a}(t=0)$ being the initial (excited) state and $e^{H\left(t\right)}$ the matrix exponential. The matrix exponential can be expressed in closed form by diagonalizing $H=Q\Lambda Q^{-1}$, with $\Lambda_{mm'}=\sigma_m\delta_{mm'}$ the diagonal matrix of complex eigenvalues $\sigma_m=i\omega_m-\gamma_m$:
\begin{equation}
\label{exp00}
e^{Ht}=Qe^{\Lambda t}Q^{-1},
\end{equation}
with $\left[e^{\Lambda t}\right]_{mm'}=e^{\sigma_{m}t}\delta_{mm'}$. If we launch an impulsive source $\mathbf{s}_{h+}(t)=\delta(t)\delta_{hh'}$ on a single scattering channel $h'$, by substituting Eqs. \eqref{exp00} into \eqref{exp01} we obtain the mode solution $a_m(t)$ at $t>0$:
\begin{equation}
    a_m(t)=\sum_{m'}\left[Qe^{\Lambda t}Q^{-1}\right]_{mm'}a_{m'}(0)=\sum_{m'}\alpha_{mm'}e^{(i\omega_{m'}-\gamma_{m'})t},
\end{equation}
expressed as the sum of complex damped exponential with $\alpha_{mm'}$ constant coefficients arising from the matrix product of $\left[Qe^{\Lambda t}Q^{-1}\right]_{mm'}a_{m'}(0)$. The Fourier transform of the mode $a_m(\omega)$ is a complex rational function:
\begin{equation}
    \tilde a_m(\omega)=\frac{1}{\sqrt{2\pi}}\sum_{m'}\frac{\alpha_{mm'}}{i(\omega-\omega_{m'})+\gamma_{m'}},
\end{equation}
with poles $\omega_{m'}+i\gamma_{m'}$. The corresponding DOS is also a rational function:
\begin{equation}
    \label{DOSfit}
    \mathrm{DOS}=\sum_m|\tilde a_m(\omega)|^2=\frac{\sum_m c_{m}\omega^m}{\prod_m \left(\omega-s_m\right)},
\end{equation}
with poles $s_m=\omega_m+i\frac{1}{\tau_m}$. To extract the Q factor, we proceed as follows. The time varying amplitude $a_m(t)$ of the electromagnetic field oscillating at frequency $\omega_m$ and decaying constant $\tau_m=\frac{2Q}{\omega_m}$ of an internal mode is $a_{m}(t)=a_0e^{i(\omega_m t-\frac{t}{\tau_m})}$, and generates a contribution to the DOS equal to:
\begin{align}
    \label{q0}
    |\tilde{a}_m(\omega)|^2=\frac{a_0}{\frac{1}{\tau_m}+i(\omega-\omega_m)}\cdot\frac{a_0^*}{\frac{1}{\tau_m}-i(\omega-\omega_m)}=\frac{|a_0|^2\tau_m}{2i}\left(\frac{1}{\omega-s_m}-\frac{1}{\omega-s_m^*}\right),
\end{align}
with $s_m=\omega_m+i\frac{1}{\tau_m}$. By equating Eqs. \eqref{DOSfit}-\eqref{q0} the quality factor $Q_m$ associated to the resonant mode at $\omega_m$ is:
\begin{equation}
    \label{q1}
    Q_m=\left\lvert\frac{\mathcal{R}\{s_m\}}{2\cdot\mathcal{I}\{s_m\}}\right\rvert,
\end{equation}
with $\mathcal{R}$ and $\mathcal{I}$ the real and imaginary part of $s_m$, respectively. The calculation of the network of mode quantities $Q_m$, $\omega_m$ and $\tau_m$ can be accomplished via a single FDTD simulation, by first calculating the DOS following the procedure outlined in the previous section and by then extracting the poles via rational fitting through Eq. \eqref{DOSfit}. For this task, we used the stable pole extraction algorithm recently developed and detailed in~\cite{Ito2018}, which is mathematically exact for rational models and can automatically detect the order of the rational polynomial in the DOS from its singular matrix.\\
Figure \ref{fitex} illustrates the accuracy of this technique in the example case of $m=1,..,7$ overlapping resonances $s_m$, with random frequencies and damping factors contained in a narrow band and generating a single apparent resonance line in the DOS (Fig. \ref{fitex}a). The solid markers in Fig. \ref{fitex}b show the position of the resonances and damping in a two dimensional $(\omega,\gamma)$ space, with the area of each marker being proportional to the $Q$ factor of each mode. Figure \ref{fitex}c presents the results of the iterative algorithm for automatic detection of the polynomial order in \eqref{DOSfit}. The efficiency of the algorithm increases exponentially and after a few iterations the system can correctly detect all the resonances (Fig. \ref{fitex}b, cross markers) composing the DOS (Fig. \ref{fitex}a, solid line), with differences between $10^{-10}$ and $10^{-25}$ (Fig. \ref{fitex}d).

\subsection{Complete representation of resonant modes}
Once the mode evolutions are obtained and stored in the $a_m(t)$ coefficients, it is possible to obtain the expression of each resonant mode from Eqs. \eqref{Eq.11} in both space $\mathbf{x}$, time $t$ and frequency $\omega$ after transforming the mode amplitudes $a_m(t)$ in the spectral domain. The main advantage of this approach lies in the fact that the quantities $a_m(t)$ are computed from a single FDTD simulation with the same setting used for the calculation of the DOS, and the complete spectrum of modes is directly available from the DOS via Eq. \eqref{dos11}.

\section*{Examples of applications}

\subsection{One dimensional structures.}
\begin{figure*}
  \centerline{\includegraphics[width=1\textwidth]{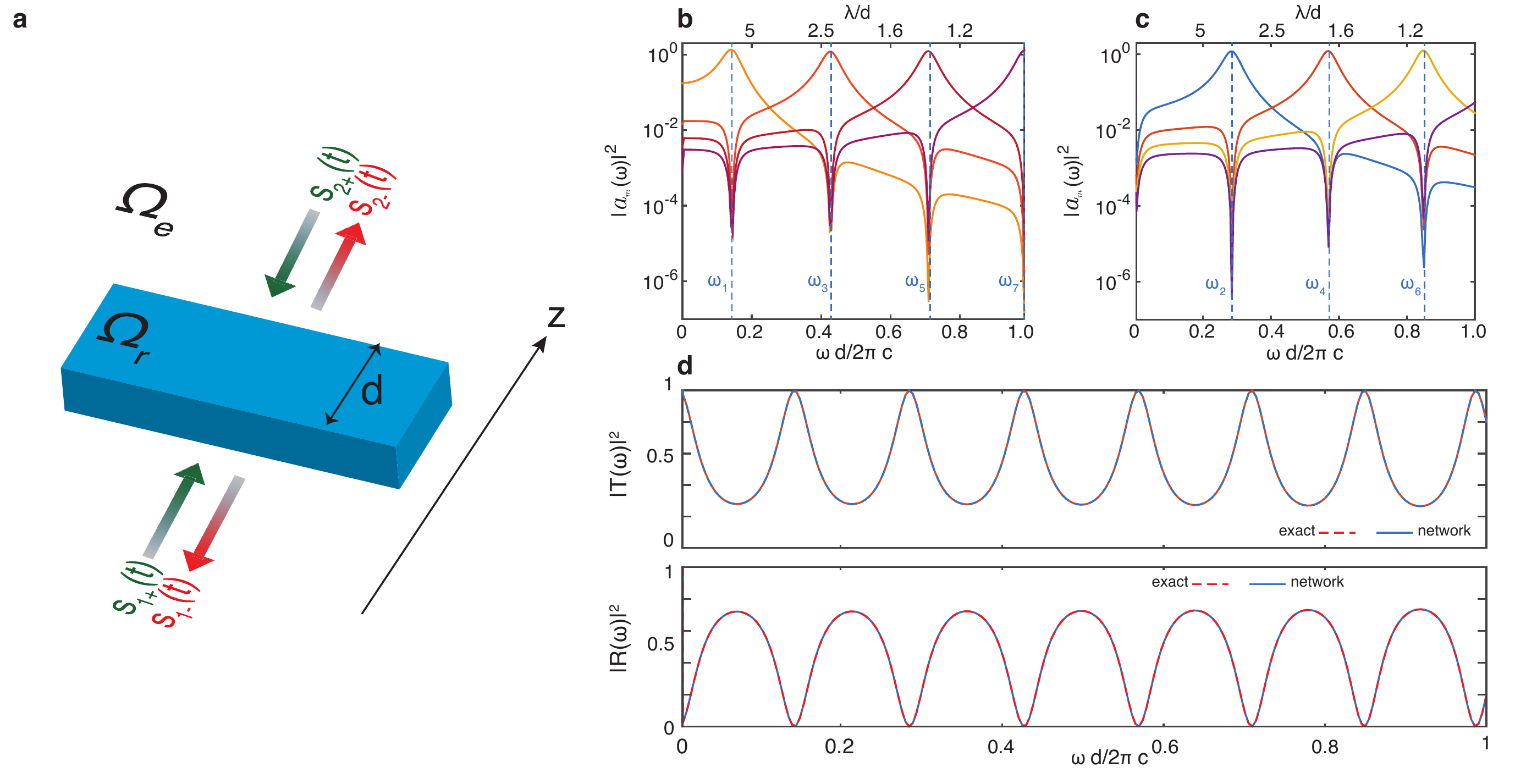}}
  \caption{\label{Panel3}
   Example of application of STCMT to a transmission and reflection problem with a dielectric cuboid slab. Panel a shows the geometry of the problem. Panels b-c illustrate the power density spectra of the resonator modes \(\left|\tilde{a}_{l}(\omega)\right|^{2}\) calculated from FDTD simulations. The analytic resonant frequencies $\omega_l$ calculated from coupled mode equations  are shown as dashed lines in the plots. Panel d shows the reflection $R$ and transmission $T$ as calculated from analytic theory and from the coupled mode approach with no fitting parameters.
   }
\end{figure*}

\begin{figure*}
  \centerline{\includegraphics[width=1\textwidth]{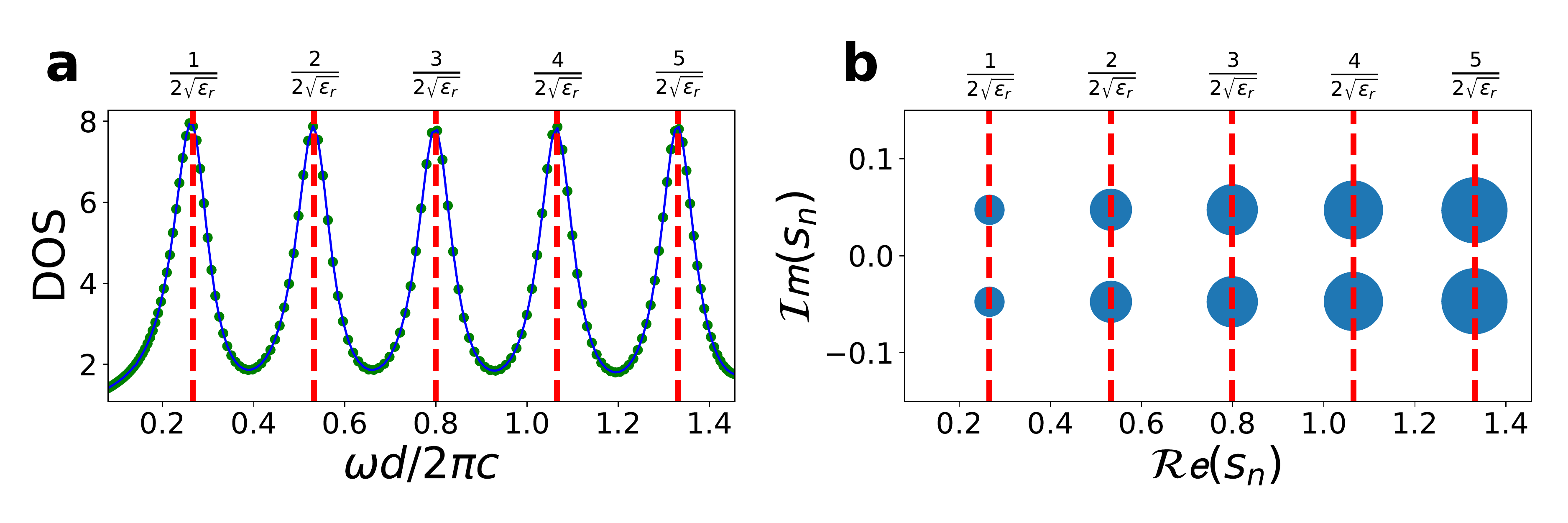}}
  \caption{\label{addon} Network representation and mode quality factor for the slab geometry of Fig. \ref{Panel3}. Panel a shows the numerically computed DOS (circle markers) and the predicted DOS ($\mathrm{DOS_P}$) from the stable rational fitting algorithm. Panel b illustrates the extracted poles positions. The sizes of points are proportional to the quality factor $Q_m$ associated with the mode. Red dotted lines represent theoretical position of resonant frequencies $\omega_m = \frac{m \pi c}{d \sqrt{\epsilon_r}}$.}
\end{figure*}

We begin by considering one dimensional structures, which illustrate the application of the theory via fully analytic calculations. Figure \ref{Panel3} shows the structure setup. The resonator region $\Omega_1$ (Fig. \ref{Panel3}a, blue region) is composed of a cuboid with thickness d along the propagation axis $z$, and with infinite sides along $x$ and $y$. The resonator space is filled with a uniform dielectric material of refractive index $n=3.5$. Despite its simplicity, this structure is sufficiently general to allow a detailed discussion of many important properties of photonics networks.\\
The system of Fig. \ref{Panel3}a has two scattering channels: when only source $s_{1+}$ is active, the reflection $R$ is measured in $s_{1-}$ and the transmission $T$ in $s_{2-}$. Conversely, when source $s_{2+}$ is launched on the structure, its reflection $R$ is retrieved in $s_{2-}$ and the transmission $T$ in $s_{1-}$. As the dielectric slab is symmetric along $z$, only one case ($s_{1+}$ or $s_{1-}$ active) is sufficient to calculate the material response.\\
Following Eqs. \eqref{Eq.14}, the frequencies of the internal modes are $\omega_m = \frac{c\pi m}{nd}$, and the spatial distribution of the magnetic modes reduce to $H_{rl}=A_0 \sin\left(\frac{l \pi z}{d}\right)$ polarized either along $x$ or $y$. External
  modes, conversely, are represented by incoming and outgoing plane waves $e^{\pm ikz}$:
\begin{align}
|z|<d/2:\label{appl.1}
&\begin{cases}
\omega_l=\frac{c\pi l}{nd},\\
H_{r}=\sum_l\frac{a(t)}{2}H_{rl}+c.c.
\end{cases},
&|z|>d/2:\begin{cases}
H_e=\tl{s}_{1+}e^{ikz}+\tl{s}_{1-}e^{-ikz},\;\;z<-d/2,\\
H_e=\tl{s}_{2+}e^{-ikz}+\tl{s}_{2-}e^{ikz},\;\;z>d/2
\end{cases}.
\end{align}
We then express the scattering matrix for the $2\times 2$ system of Fig. \ref{Panel3}:
\begin{equation}\label{appl.2}
\left[\begin{array}{c}{\tilde{s}_{1-}} \\ {\tilde{s}_{2-}}\end{array}\right]=\left[\begin{array}{cc}{\tilde{C}_{11}} & {\tilde{C}_{12}} \\ {\tilde{C}_{12}} & {\tilde{C}_{22}}\end{array}\right] \cdot\left[\begin{array}{c}{\tilde{s}_{1+}} \\ {\tilde{s}_{2+}}\end{array}\right], \quad \qquad \tilde{C}=\left[\begin{array}{cc}{-1} & {0} \\ {0} & {-1}\end{array}\right]
\end{equation}
in which $\tilde{C}_{12}=\tilde{C}_{21}=0$ and $\tilde{C}_{11}=\tilde{C}_{22}=-1$ arise from PEC boundary condition at the resonator space $\Omega_n$, originating reflections for each incoming source when light is injected from the external space $\Omega_e$. In the geometry of Fig. \ref{Panel3}a, the scattering matrix is already in diagonal form.\\ 
By exploiting the self consistency relations \eqref{Eq.22}, we can express the diagonal elements of the damping matrix $\Gamma$ from the coupling coefficients $K_{ij}=|K_{ij}| e^{i\theta_{ij}}$:
\begin{equation}\label{appl.3}
\Gamma_{i i}=\frac{\left|K_{i 1}\right|^{2}}{2}+\frac{\left|K_{i 2}\right|^{2}}{2}
\end{equation}
The symmetry of the resonator structure along $z$ implies that damping factors along channels 1 and 2 are the same: \(\left|K_{i 1}\right|=\left|K_{i 2}\right|=\sqrt{\Gamma_{i i}}\). The remaining elements of the damping matrix are then:
\begin{equation}\label{appl.4}
\Gamma_{i j}=\frac{K_{i 1} K_{j 1}^{*}}{2}+\frac{K_{i 2} K_{j 2}^{*}}{2}=\frac{\sqrt{\Gamma_{i i} \Gamma_{j j}}}{2}\left[e^{i\left(\theta_{i 1}-\theta_{j 1}\right)}+e^{i\left(\theta_{i 2}-\theta_{j 2}\right)}\right]
\end{equation}
In Eq. \eqref{appl.4} two cases are possible due to the $z$ symmetry of the system, and each internal
mode can either decay symmetrically or anti-symmetrically in the scattering channels:
\begin{equation}\label{appl.5}
\left\{\begin{array}{l}{\theta_{i 1}=\theta_{i 2}+2 m \pi, \text { symmetric }} \\ {\theta_{i 1}=\theta_{i 2}+(2 m+1) \pi, \text { anti-symmetric }}\end{array}\right.
\end{equation}
These relations imply that when internal modes $i$, $j$ possess opposite symmetry along $z$, we have \(\theta_{i 1}-\theta_{j 1}=- \left(\theta_{i 2}-\theta_{j 2}\right)\) and  $\Gamma_{ij} = 0$. Conversely, when modes $i$, $j$ have the same symmetry \(\theta_{i 1}-\theta_{j 1}=\left(\theta_{i 2}-\theta_{j 2}\right)\) and \(\Gamma_{i j}=\pm \sqrt{\Gamma_{i i} \Gamma_{j j}}\), with the plus sign if both modes are even and with minus sign if modes are odd. These cases are summarized as follows:
\begin{equation}\label{appl.6}
\Gamma_{i j}=\left\{\begin{array}{ll}{0,} & {\text { modes $i$ and $j$ have opposite symmetry on $z$ }} \\ { \pm \sqrt{\Gamma_{i i} \Gamma_{j j}},} & {\text { $i$ and $j$ have same symmetry }}\end{array}\right.
\end{equation}
By substituting the expression of $\Gamma_{ij}$ in the coupled mode equations \eqref{exp01}, we find the following main result, which holds for all modes having the same parity:
\begin{equation}\label{appl.7}
\tilde{a}_{m}\left(\omega_{l}\right)=\left\{\begin{array}{ll}{0,} & {m \neq l} \\ {\frac{1}{\sqrt{\Gamma_{l l}}},} & {m=l}\end{array}\right.
\end{equation}
This equation states that the amplitude of each internal mode \(\tilde{a}_{m}(\omega_m)\), evaluated at the resonant frequency \(\omega_{m}\)of the mode, furnishes the amplitude of the damping factor $\Gamma_{ll}$, and goes to zero at all the resonance frequencies of the other modes. Equation \eqref{appl.7} is sufficient to calculate all the elements of the damping matrix via Eq. \eqref{appl.3} and \eqref{appl.6}.  Equation \eqref{appl.7} is a direct consequence of the phase matching condition discussed for the coupling matrix coefficients expressed by Eq. \eqref{Eq.28}.\\
We verified these results, and in particular the validity of \eqref{appl.7}, by FDTD simulations. We begin by calculating the coefficients \(a_{n}(t)\) by projecting over the magnetic field eigenmodes defined by \eqref{appl.1}. Figure \ref{Panel3} (b-c) shows the spectral distribution \(\left|\tilde{a}_{m}(\omega)\right|^{2}\) of the first seven modes in the expansion. The modes are grouped into even (Fig. \ref{Panel3}b) and odd (Fig. \ref{Panel3}c). In agreement with the analytic results based on \eqref{appl.7}, the amplitude of each mode vanishes at the internal resonant frequency \(\omega_{l \neq m}\) of the other modes with the same symmetry along $z$.\\
From the amplitude intensity at the mode internal frequency \(\left|\tilde{a}_{l}(\omega)\right|^{2}\), we apply \eqref{appl.7} and calculate all the elements of the coupling $K$ and damping $\Gamma$ matrix. Fig. \ref{Panel3}d illustrates the transmissivity and reflectivity of the structure calculated from analytic solution via multilayer theory (dashed line) and from the network model based on Eqs. \eqref{Eq.23}. The solutions are exactly the same, with relative differences below $10^{-16}$. In the analysis we did not use any fitting curve, background or parameters, but calculated all coefficients from the analysis of internal modes by using Eq. \eqref{appl.7}.\\
Figure \ref{addon} shows the representation of the network mode parameters extracted from the poles of the DOS (a, circle markers) of the resonator of Fig. \ref{Panel3}a. The DOS is calculated from Eq. \eqref{newdos} by summing up the spectral mode densities $|\tilde{a}_m(\omega)|$ as illustrated in Fig. \ref{Panel3}b-c. The rational model representing the DOS via Eqs. \eqref{DOSfit} is reported as a solid line in Fig. \ref{addon}a. The network representation illustrated in Fig. \ref{addon}b correctly predicts a mode network composed by modes with identical damping factors $\gamma_m$ and resonant frequencies following the exact analytic formula $\frac{\omega_md}{2\pi c}=\frac{m}{2\sqrt{\epsilon_r}}$ (red dashed line) of a cuboid resonator terminated by PMC boundary conditions. The corresponding quality factor of the modes (Fig. \ref{addon}b, solid area of each marker) increases in $\omega$ due to the increasing resonant frequency $\omega_m$ of each mode.  

\subsection{Two and three dimensional structures}
\begin{figure*}
  \centerline{\includegraphics[width=1\textwidth]{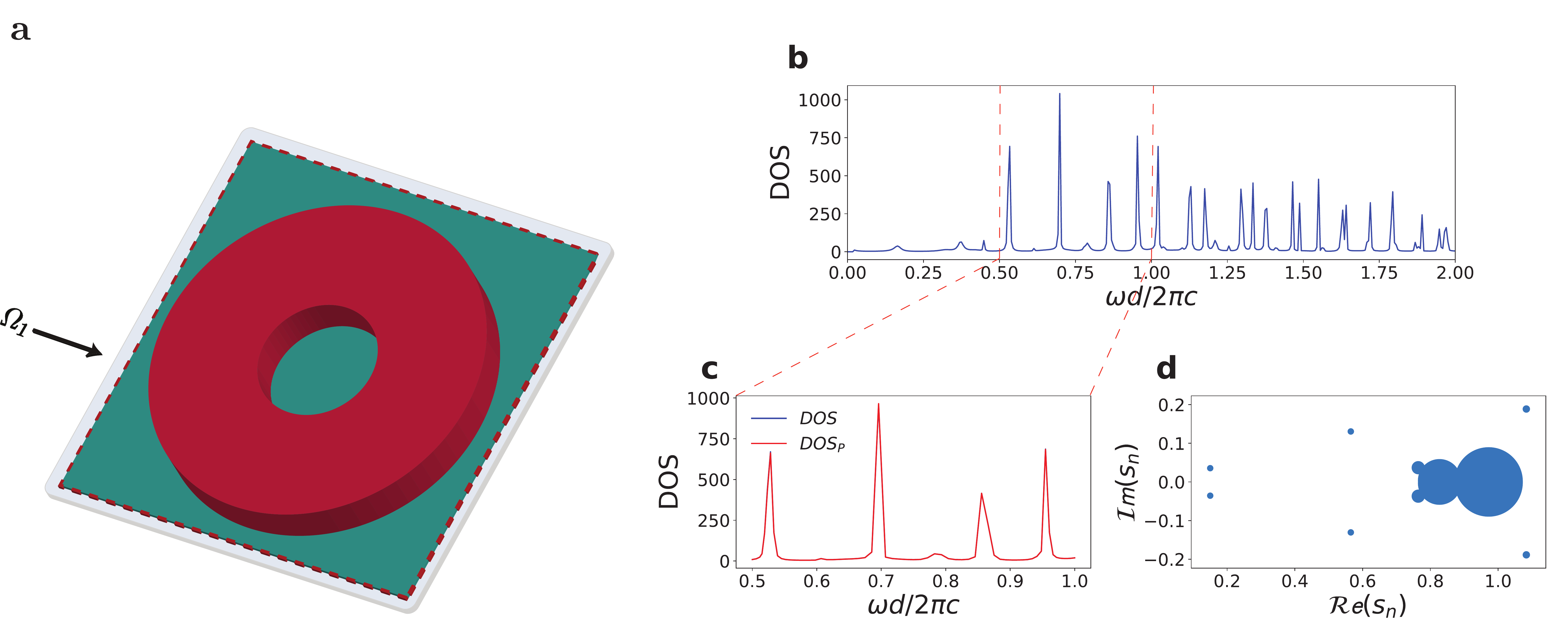}}
  \caption{\label{Panel4}
    Analysis of a two dimensional non periodic ring resonator.  Panel a schematically represents the resonator structure (red) with the cuboid volume region $\Omega_n$ (green) denoting the cavity space.
    (b) Corresponding DOS. (c) Zoomed DOS in the visible range. (d) Poles map, with each pole represented as a circle with size proportional to the mode quality factors $Q_m$.
   }
\end{figure*}

\begin{figure*}
  \centerline{\includegraphics[width=1\textwidth]{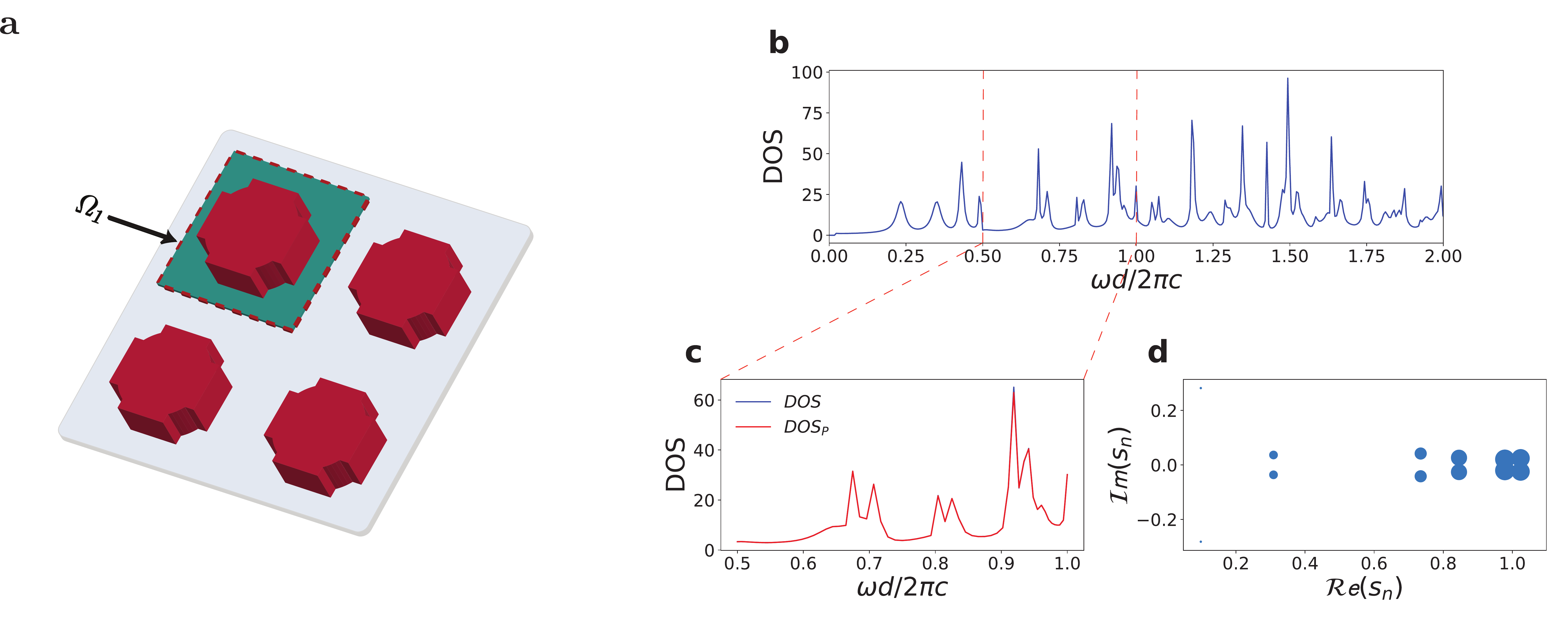}}
  \caption{\label{Panel5}
    Periodic resonator configuration. (a) resonator array (red) and resonator space region $\Omega_1$ (green). (b) Zoom of the DOS into the visible spectrum region (c). (d) Poles map. The sizes of points are proportional to the quality factor $Q_m$ associated with the mode.
   }
\end{figure*}

\begin{figure*}
  \centerline{\includegraphics[width=1\textwidth]{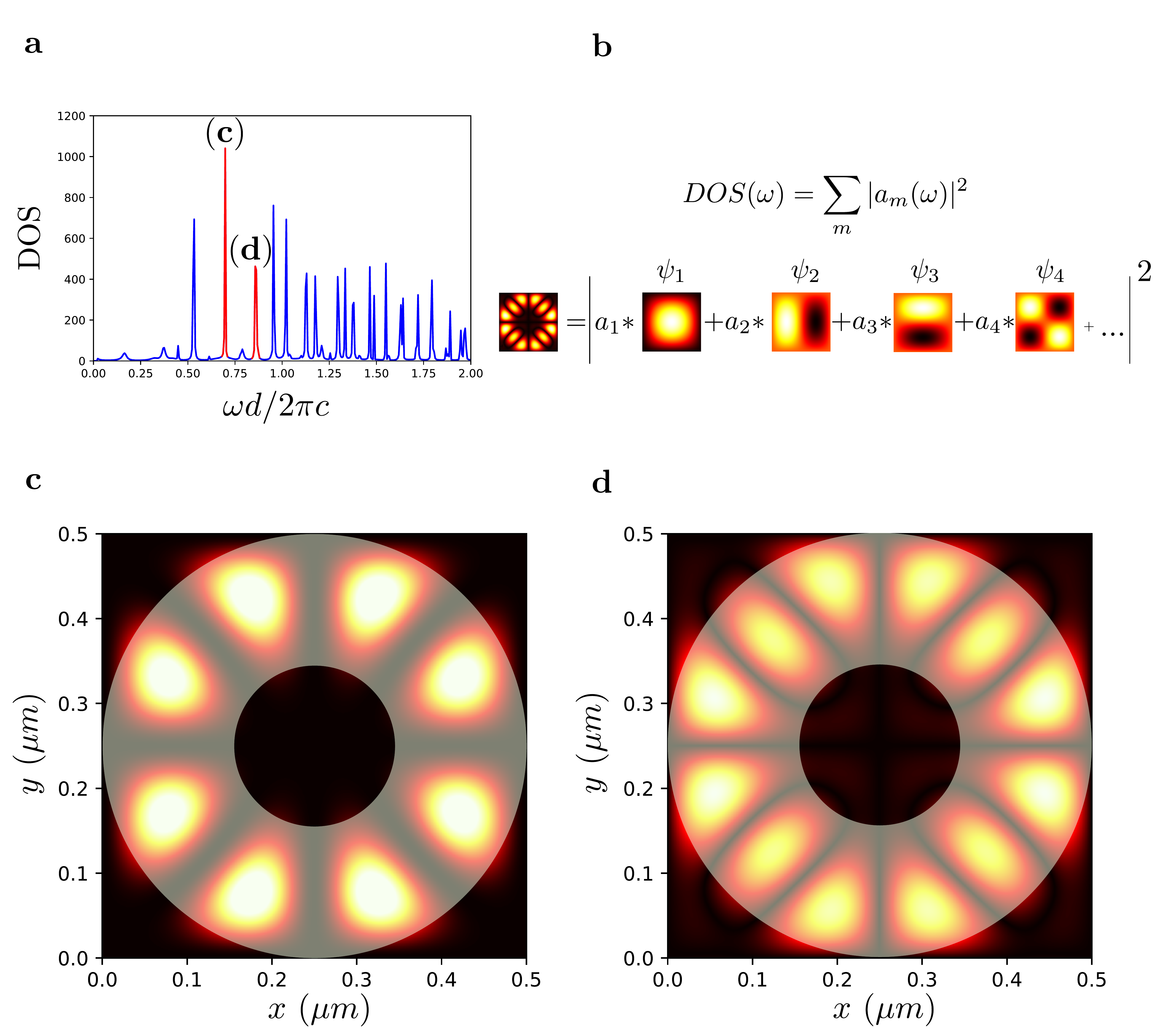}}
  \caption{\label{Panel6}
    Spatial modes representations for two different frequencies (b and c points in panel a) of the DOS of the ring resonator of Fig. \ref{Panel4}. (b) Visual illustration of the mode decomposition via orthogonal modes of the resonator space. (c-d) Spatial distributions of the corresponding electromagnetic modes. The superimposed transparent image in (c-d) shows the ring resonator refractive index distribution.
   }
\end{figure*}

\begin{figure*}
  \centerline{\includegraphics[width=1\textwidth]{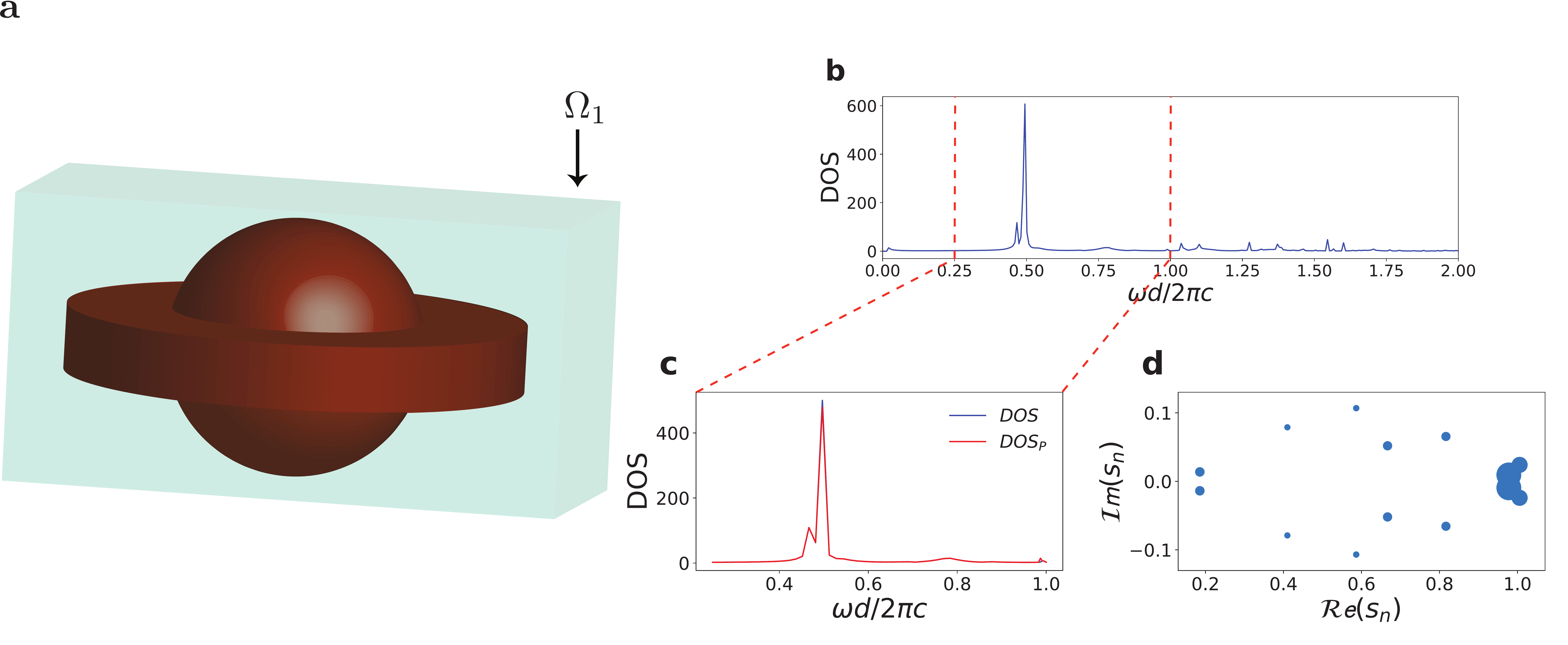}}
  \caption{\label{Panel7}
   Analysis of three dimensional structure. (a) Resonator consisting of a superposition of a sphere and a disk (red) with a cubic resonator space $\Omega_1$ (green). (b) DOS and zoom into the visible spectrum region (c). (d) Poles map. The sizes of points are proportional to the quality factor $Q_m$ associated with the mode.}
\end{figure*}

We apply the STCMT approach to describe more complicated geometries for both non-periodic (Fig. \ref{Panel4}) and periodic (Fig. \ref{Panel5}) boundary conditions. 
Figure \ref{Panel4}a illustrates the schematic representation of a two dimensional resonator ring shaped geometry with 0.15 $\mu m$ and 0.5 $\mu m$ sizes for the inner and outer diameters, respectively. We define a single cuboid resonator space $\Omega_1$ (Fig. \ref{Panel4}a green area) that includes the whole resonator, and use the orthogonal mode set of the cuboid volume without the resonator space, as employed in Fig. \ref{Panel2}. The corresponding calculated DOS is fitted with the rational expression \eqref{DOSfit} on Fig. \ref{Panel4}b. Figure \ref{Panel4}c provides a zoom of the results on a target frequency range covering most of a visible spectra. The red dotted lines represent the boundaries of the target frequency range. The  resonance network with the various $Q_m$ factors of the resonator modes is presented in Fig. \ref{Panel4}c.\\   
Figure \ref{Panel5}a-d shows equivalent results obtained from a periodic two dimensional resonator characterized by a complex geometry consisting of the concentric superposition of a cross shaped geometry and a disk. In this case, due to stronger mode coupling via periodic neighbor resonators, the peaks in DOS spectra are wider and the overall DOS is smoother.\\
As discussed in the theory, the STCMT allows to obtain the full dynamics of the field, including the spatial distribution of any resonant mode existing inside the resonator. We illustrate this approach with reference to the non periodic ring resonator of Fig. \ref{Panel4}a. We consider two resonant peaks in the DOS, labeled (c) and (d) in Fig. \ref{Panel6}a, and extract the modes spatial profile by Eq. \eqref{Eq.11} by using the orthogonal eigenfunctions Eqs. \ref{Eq.13} (Fig. \ref{Panel6}b) of the cuboid resonator space $\Omega_1$ used in the projections of the time varying coefficients $a_m(t)$.\\
The corresponding energy distributions of the resonant modes are portrayed in Fig. \ref{Panel6}c-d. They represent whispering-gallery modes of the ring-like geometry.
Figure \ref{Panel7} shows the calculation results for a three dimensional resonator structure, delimited by a cubical $\Omega_1$ resonator space. The resonator here is a compound shape consisting of a concentric superposition of a sphere and a disk. The mode network of this structure (Fig. \ref{Panel7}d) is mainly represented by two close resonances with quality factors $Q_1=52$ and $Q_2=21$.\\   

\section*{Conclusion}
In this work we formulate an exact spatio-temporal coupled mode theory for arbitrary resonator structures, derived with orthogonal and complete eigenmodes obtained from Maxwell equations with generalized operators. Using this theory, it is possible to provide an exact representation of the electromagnetic dynamics in both space, time and frequency via a simple set of exact equations of motion, in which all relevant quantities such as the DOS, the modes quality factors, and the modes spatial distribution can be calculated numerically from a single first principle simulation. We provide examples of this approach in one, two and three dimensional optical structures. We believe that this approach can help the design of photonics systems based on complex multi mode interactions, providing an exact formulation of coupled mode equations in general conditions of overlapping resonances and for arbitrarily defined materials and resonator geometries. 

\bibliography{gen_max.bib}

\section*{Acknowledgements}
The authors acknowledge support from KAUST (OSR-2016-CRG5-2995) and Shaheen supercomputer from the Kaust Supercomputing Laboratory (KSL). 

\section*{Competing Interest}
The authors declare no competing interests.


\clearpage

\clearpage

\clearpage

\clearpage

\clearpage

\clearpage

\clearpage

\clearpage

\clearpage

\end{document}